\documentclass[conference]{IEEEtran}
\IEEEoverridecommandlockouts
\usepackage{url} 
\usepackage{xspace} 
\usepackage{comment} 

\usepackage{cite}
\usepackage{amsmath,amssymb,amsfonts}
\usepackage{algorithm}
\usepackage{algpseudocode}
\usepackage{enumitem}
\usepackage{graphicx}
\usepackage{caption,subcaption}
\usepackage{textcomp}
\usepackage{xcolor}

\def\BibTeX{{\rm B\kern-.05em{\sc i\kern-.025em b}\kern-.08em
    T\kern-.1667em\lower.7ex\hbox{E}\kern-.125emX}}

\newtheorem{definition}{Definition}

\newtheorem{example}{Example}

\newcounter{experiment}

\begin{document}

\title{Dataset Discovery in Data Lakes}
 
\author{\IEEEauthorblockN{Alex Bogatu, Alvaro A.A. Fernandes, Norman W. Paton, Nikolaos Konstantinou}
\IEEEauthorblockA{\textit{School of Computer Science,} 
\textit{University of Manchester,}
Manchester, UK \\
alex.bogatu@manchester.ac.uk}
}

\maketitle

\begin{abstract}
Data analytics stands to benefit from the increasing availability of datasets that are held without their conceptual relationships being explicitly known. When collected, these datasets form a data lake from which, by processes like data wrangling, specific target datasets can be constructed that enable value--adding analytics. Given the potential vastness of such data lakes, the issue arises of how to pull out of the lake those datasets that might contribute to wrangling out a given target. We refer to this as the problem of dataset discovery in data lakes and this paper contributes an effective and efficient solution to it. Our approach uses features of the values in a dataset to construct hash--based indexes that map those features into a uniform distance space. This makes it possible to define similarity distances between features and to take those distances as measurements of relatedness w.r.t. a target table. Given the latter (and exemplar tuples), our approach returns the most related tables in the lake. We provide a detailed description of the approach and report on empirical results for two forms of relatedness (unionability and joinability) comparing them with prior work, where pertinent, and showing significant improvements in all of precision, recall, target coverage, indexing and discovery times.     
\end{abstract}


\section{Introduction}
\label{sec:intro}

The number of external (and internal) datasets, of increasing diversity, that are available for organizations to use continues to grow, and we find ourselves past the point where one could impose upon any collection of such datasets any global, conceptually cohesive, model that captures their interrelationships. It has become easy to amass datasets that have great potential for analytics, but with the lack of conceptual cohesion comes a greater difficulty to even discover the most useful datasets for say, a given analytic task. 

There is still no consensus on what the notion of a data lake denotes. In this paper, we take a \emph{data lake} to be a repository whose items are datasets about which, we assume, we have no more metadata than, when in tabular form, their attribute names, and possibly their domain-independent types (i.e., string, integer, etc.). We view open government data repositories as exemplar data lakes. 

We view the process of doing analysis on data from a data lake as being dependent on data wrangling \cite{Furche-16} and, as such, comprising many stages (e.g., \cite{Koehler-17, Konstantinou-17}). This paper views the basic, initial stage as one of \emph{dataset discovery}, that filters the (otherwise unmanageable) input for subsequent stages such as schema matching (e.g., \cite{Rahm-01}), format transformation (e.g., \cite{Bogatu-19}), or schema mapping generation (e.g., \cite{Mecca-09, Mazilu19}), i.e., given a target (which ideally includes exemplar tuples and expected attribute names), find which datasets are most useful as inputs for data wrangling. By most useful, we mean datasets (or possibly projections thereof) that are \emph{unionable} with the target, and, desirably, \emph{joinable} with each other. 
Given a target, our objective is to identify related datasets from a data lake that are relevant for populating as many target attributes as possible.

\begin{example}
Consider Figure \ref{fig:runex}. The target $T$ contains information about general practices (i.e., family doctors/primary care centers). We want to find tables (e.g., $S_1$ and $S_2$) useful for populating $T$. Moreover, assuming $S_3$ and $T$ are not strongly related, we want to find join opportunities (e.g., of \emph{Practice Name/Practice} in $S_1/S_2$ with \emph{GP} in $S_3$) that allow us to increase target coverage by populating  \emph{Hours} in $T$.
\label{ex:intro}
\end{example}

\begin{figure*}[t]
\begin{center}
\small

\begin{tabular}{|c|c|c|c|c|}
\multicolumn{5}{l}{$S_1$: Source: GP practices} \\
\hline
Practice Name & Address & City & Postcode & Patients \\
\hline
\hline
Dr E Cullen & 51 Botanic Av & Belfast & BT7 1JL & 1202 \\
\hline
Blackfriars & 1a Chapel St & Salford & M3 6AF & 3572 \\
\hline
\end{tabular}
~
\begin{tabular}{|c|c|c|c|}
\multicolumn{4}{l}{$S_2$: Source: GP funding} \\
\hline
Practice & City & Postcode & Payment \\
\hline
\hline
The London Clinic & London & W1G 6BW & 73648 \\
\hline
Blackfriars & Salford & M3 6AF & 15530 \\
\hline
\end{tabular}

\medskip

\begin{tabular}{|c|c|c|}
\multicolumn{3}{l}{$S_3$: Source: Local GPs} \\
\hline
GP & Location & Opening hours \\
\hline
\hline
Blackfriars & Salford & 08:00-18:00 \\
\hline
Radclife Care & - & 07:00-20:00 \\
\hline
\end{tabular}
~
\begin{tabular}{|c|c|c|c|c|}
\multicolumn{5}{l}{$T$: Target: GPs} \\
\hline
Practice & Street & City & Postcode & Hours \\
\hline
\hline
Radclife & 69 Church St & Manchester & M26 2SP & 07:00-20:00 \\
\hline
Bolton Medical & 21 Rupert St & Bolton & BL3 6PY & 08:00-16:00 \\
\hline
\end{tabular}

\normalsize
\end{center}
\caption{Example Tables}
\label{fig:runex}
\end{figure*}

In this paper, we contribute a solution to the data discovery problem. Our approach, which we refer to as $D^3L$ (for Dataset Discovery in Data Lakes), can broadly be seen as similarity-based in the sense that, from the attribute names and values in each dataset in the lake, we extract features that convey signals of similarity. We extract five types of features that we map to values in \textit{locality-sensitive hashing} (LSH) indexes \cite{Indyk-1998}, thereby guaranteeing that shared bucket membership is indicative of similarity as per the hash function used. Specifically, we make the following contributions:
\begin{itemize}[leftmargin=*]
    \item We propose a new distance--based framework that, given a target, can efficiently determine the relatedness between target attributes and attributes of datasets in a lake. We do this using five types of evidence: (i) \emph{attribute name similarity}, when schema--level information is available; (ii) \emph{attribute extent overlap}, when attributes share common values; (iii) \emph{word--embedding similarity}, when attributes are semantically similar but have different value domains; (iv) \emph{format representation similarity}, when attribute values follow regular representation patterns; and (v) \emph{domain distribution similarity}, for numerical attributes.
    \item We show how to map the signal from each of the above evidence types onto a common space such that the resulting attribute distance vectors combine the separate measurements of relatedness, and propose a weighting scheme that reflects the signal strength from different evidence types.
    \item We extend the notion of relatedness to tables whose similarity signal with the target is weak but that join with tables that contribute values to additional target attributes.
    \item We empirically show, using both real-world and synthetic data, that $D^3L$ is significantly more effective and more efficient than the state--of--the--art (specifically, \cite{Fernandez-18,Nargesian-2018}).
\end{itemize}

\section{Related Work} \label{sec:relwork}

Data lakes are usually seen as vast repositories of company, government or Web data (e.g., \cite{Cafarella-08,Elmeleegy-09}). Previous work has considered dataset discovery in the guise of table augmentation and stitching (e.g., \cite{Lehmberg-17,Ling-13}), unionability discovery, or joinability discovery (e.g., \cite{DasSarma-2012,Nargesian-2018,Fernandez-18,Fernandez18_sem}). We add to this work with a focus on a notion of relatedness, defined in the next section, construed as unionability and/or joinability.

\smallskip \noindent \textbf{LSH}. We build upon \textit{locality--sensitive hashing} (LSH), an approach to nearest--neighbours search in high--dimensional spaces  \cite{Indyk-1998}. LSH requires hash functions whose collision probability is high for similar inputs, and lower for those that are more different. Several such functions have been proposed for different similarity metrics, e.g.,\cite{Datar-04, Broder-1997, Charikar-2002}. Given an LSH index, the similarity degree of two items is given by the number of buckets, i.e., index entries, containing both items, and the kind of similarity achieved depends on the hash function used. In this paper we rely on two such hash functions that return hash values with high probability of collision for inputs with high Jaccard similarity: \textit{MinHash} \cite{Broder-1997}, and with high cosine similarity: \textit{random projections} \cite{Charikar-2002}. 

In practice, we use LSH Forest \cite{Bawa-05}, an extension to LSH that largely ensures that for an answer size $k$, the search time varies little with the size of the repository. One other LSH improvement, compatible with our use case, is LSH Ensemble \cite{Zhu-2016}, which proposes an indexing scheme that aims to overcome the weaknesses of \textit{MinHash} when used on sets with skewed lengths. 

\smallskip \noindent \textbf{LSH--based dataset discovery}. LSH has been adopted by the data management research community due to its useful properties regarding similarity estimation, associated with linear retrieval times w.r.t the search space size \cite{Miller-18}. One example is Aurum \cite{Fernandez-18} (and its extension from \cite{Fernandez18_sem}), a system to build, maintain and query an abstraction of a data lake as a knowledge graph. Similarly, Table Union Search \cite{Nargesian-2018} focuses on the problem of unionability discovery between datasets, treated as an LSH index lookup task. As we do, both proposals use LSH--based indexes to efficiently search for related attributes in data repositories. While the underlying data structures used in both cases are similar to the ones we rely on, there are a number of key differences: (i) we make use of more types of similarity, whose combined import is to inform decisions on relatedness with a diversity of signals; (ii) we adopt an approach based on schema-- and instance--level fine--grained features that prove more effective in identifying relatedness, especially in cases when similar entities are inconsistently represented; (iii) we map these features to a uniform distance space that offers a holistic view on the notion of relatedness between attributes, to which each type of similarity evidence contributes, as instructed by an underlying weighting scheme.

\smallskip \noindent \textbf{Web data integration}: The discovery of unionable/joinable Web tables has been studied in Octopus \cite{Cafarella-09} which combines search, extraction and cleaning operators to create clusters of unionable tables by means of string similarities and Web metadata. Das Sarma \textit{et al.} \cite{DasSarma-2012} identify entity complementary (unionable) and schema complementary (joinable) tables by using knowledge--bases to label datasets at instance and schema levels, leading to a decision on their unionability and joinability. We too search for such tables but, because we envisage the need for downstream wrangling, we assume a target table and refrain from relying on Web knowledge--bases or external metadata as such data will not always be available.

\smallskip \noindent \textbf{Data lake management systems}: Data lakes have been the focus of recent research on data management systems, e.g., \cite{Terrizzano15, Halevy16}. Such proposals focus on data lifecycle and rely on extensible metadata models and parsing frameworks for different data types, tailored for the challenges faced by the organization that builds and uses the data lake, e.g., Goods \cite{Halevy16}, is highly oriented towards rapidly changing data sets. 

\section{Relatedness discovery}

Before describing our approach in detail, we formally define the \textit{relatedness} of a dataset $S$ w.r.t. a target $T$ as follows:

\begin{definition} \label{def:relatedness}
    Given a dataset $S$ with attributes $a_1, \ldots, a_n$ and a target dataset $T$ with attributes ${a'}_1, \ldots, {a'}_m$, we say that $S$ and $T$ are \textit{related} iff $\exists a_i \in \{a_1, \ldots, a_n\}$, so that $a_i$ contains values drawn from the same domain represented by some attribute ${a'}_j \in \{{a'}_1, \ldots, {a'}_m\}$, and, therefore, is relevant for populating ${a'}_j$, i.e., $a_i$ and ${a'}_j$ are \textit{attribute--level related}.
\end{definition}

Given two datasets $S_1$ and $S_2$ related w.r.t. a target $T$, the following properties follow from Definition \ref{def:relatedness}:

\begin{itemize}[leftmargin=*,align=left]
    \item $S_1$ and $S_2$ can have different degrees of relatedness to $T$, subject to how many of their attributes are related to some target attribute and to how strongly related the attributes are.
    \item $S_1$ and $S_2$ are \textit{unionable} on the attributes related to the same target attribute and each is unionable with the target itself. We focus on relatedness--by--unionability in this section.
    \item If $S_1$ and $S_2$ are \textit{joinable} as well, then the projection from their join result of the attributes related to some target attribute is, potentially, related to $T$ as well. We explore this property in Section \ref{sec:joins}.
\end{itemize}

We consider relatedness to also imply similarity, and quantify the former using distance measures: the closer, the more similar, and the more similar, the more related. 

\subsection{Attribute Relatedness: Relatedness evidence} \label{sec:attrRel}

We first aim to identify related attributes, i.e., attributes whose values can be used to populate some attribute in the target, and to quantify their degree of relatedness. Strictly, this can only be done if they store values for the same property type of the same real world entity. However, data lakes are characterized by a dearth of metadata. There is a need, then, to decide whether two attributes from two datasets are related relying only on evidence that the datasets themselves convey.

We use five types of evidence for deciding on attribute relatedness: names ($\mathbb{N}$), values ($\mathbb{V}$), formats ($\mathbb{F}$), word-embeddings ($\mathbb{E}$) \cite{Mikolov-13}, and domain distributions ($\mathbb{D}$). From names we derive $q$--grams; from values we derive tokens, format--describing regular expressions and word--embeddings; and from extents we derive domain distributions. Note that, in the case of both attribute names and attribute values, 
we break up string representations with a view to obtaining finer-grained evidence. The motivation is the expected ``dirtiness" of the data lake, e.g., attributes may have names or values that denote the same real-world entity but are represented differently. 
Using finer--grained evidence implies that our approach is lenient in identifying related attributes, reducing the impact of dirty data. This is an important point of contrast with related work, as the experimental results will show.

Let $a$ and $a'$ be attributes with extents $[\![a]\!]$ and $[\![a']\!]$, resp. We now describe how we aim to capture similarity signals for each type of evidence:

\begin{description}[leftmargin=*,align=left]
\item[$\mathbb{N}$]: given an attribute \textbf{name}, we transform it into a set of $q$-grams ($q$set, for short), aiming to construe relatedness between attribute names as the Jaccard distance between their $q$sets. Let $Q(a)$ denote the $q$set of $a$.
\item[$\mathbb{V}$]: given an attribute \textbf{value}, we transform it into a set of informative tokens ($t$set, for short). By informative, we mean a notion akin to term-frequency/inverse-document-frequency (TF/IDF) from information retrieval, as explained later. We aim to construe relatedness between attribute values as the Jaccard distance between their $t$sets. Let $T(a)$ denote the union of the $t$sets of every value  in $[\![a]\!]$.
\item[$\mathbb{F}$]: given an attribute value, we represent its \textbf{format} (i.e., the regular, predictable structure of, e.g., email addresses, URIs, dates, etc.) by a set of regular expressions ($r$sets, for short) grounded on a set of primitives we describe later. We aim to construe relatedness between attribute value formats as the Jaccard distance of their $r$sets. Let $R(a)$ denote the union of the $r$sets of every value $v$ in $[\![a]\!]$. 
\item[$\mathbb{E}$]: given an attribute value that has textual content, we capture its context-aware semantics as described by a \textbf{word--embedding} model (WEM) \cite{Grave-17}, as follows: each word in the attribute value is assigned a vector (with  WEM-specific dimension $p$) that denotes its position in the WEM-defined space. The $p$-vectors of each such word are then combined into a $p$-vector for the whole attribute. We aim to construe relatedness between attribute values with textual content as the cosine distance of their vectors. Let $\vec{a}$ denote the set of word-embedding $p$-vectors of every value in $[\![a]\!]$.
\item[$\mathbb{D}$]: given a numeric attribute, only $\mathbb{N}$ and $\mathbb{F}$ are useful in construing similarity, as the others (viz., $\mathbb{V}$ and $\mathbb{E}$) are dependent on the existence of structural components (viz., tokens and words) that can only be reasonably expected in non-numeric data. So, we aim to construe relatedness between numeric attribute values as the Kolmogorov-Smirnov statistic ($KS$) \cite{Stats-99} over their extents understood as samples of their originating \textbf{domain}. The smaller KS is, the closer the attributes are w.r.t. to their value distribution.
\end{description}

\subsection{Attribute Relatedness: Distance Computation} \label{sec:indexConstr}

Each of the five types of evidence above gives rise to a distance measure bounded by the $[0,1]$ interval. Given two attributes $a$ and $a'$, from different features of their respective names and extents, all of which carry useful but different signals of relatedness, we can compute the following distances:

\begin{description}
\item[name]: $D_{\mathbb{N}}(a,a') = 1-(Q(a) \cap Q(a'))/(Q(a) \cup Q(a'))$, i.e., the Jaccard distance between their $q$sets.
\item[value]: $D_{\mathbb{V}}(a,a') = 1-(T(a) \cap T(a'))/(T(a) \cup T(a'))$, i.e., the Jaccard distance between their $t$sets.
\item[format]: $D_{\mathbb{F}}(a,a') = 1-(R(a) \cap R(a'))/(R(a) \cup R(a'))$, i.e., the Jaccard distance between their $r$sets.
\item[embedding]: $D_{\mathbb{E}}(a,a') =  1-((\vec{a}^{T}\vec{a}')/(|\vec{a}|\cdot|\vec{a}'|))$, i.e., the cosine distance between their word-embedding vectors. 
\item[domain]: $D_{\mathbb{D}}(a,a') = KS([\![a]\!], [\![a']\!])$, 
i.e., the $KS$ computed over their extents.
\end{description}

In order to avoid carrying pairwise comparisons in computing the above distances, as others have done, e.g., \cite{Nargesian-2018}, \cite{Fernandez-18}, we adopt an approximate solution based on LSH that offers efficient distance computation, at the potential expense of accuracy. To this end, we use Jaccard and cosine distances because of their property of being \textit{locality--sensitive} (\cite{Broder-1997, Charikar-2002}). Specifically, the probability that MinHash/random--projections returns the same hash value for two sets is approximately equal to their Jaccard/cosine similarity. Since $\mathbb{N}$--, $\mathbb{V}$--, $\mathbb{F}$--relatedness are grounded on Jaccard similarity, and $\mathbb{E}$--relatedness is grounded on cosine similarity, we use MinHash/random projections to efficiently approximate the above distances. We do not use the same strategy for $\mathbb{D}$--relatedness because there is no LSH hashing scheme that leads to analogous gains.

In our approach, given a data lake $\mathcal{S}$ and a target table $T$, finding the set of $k$--most related datasets in $\mathcal{S}$ to $T$ is a computational task performed after indexing $\mathcal{S}$. For a given $\mathcal{S}$, we build four LSH indexes, which, resp., are used to compute $\mathbb{N}$--, $\mathbb{V}$--, $\mathbb{F}$--, and $\mathbb{E}$--relatedness between attributes. We call these indexes $I_\mathbb{N}$, $I_\mathbb{V}$, $I_\mathbb{F}$, and $I_\mathbb{E}$, resp. Given two attributes $a$ and $a'$, they are $\mathbb{N}$ (resp., $\mathbb{V}$, $\mathbb{F}$, and $\mathbb{E}$)--related if $a' \in I_{\mathbb{N}}.lookup(a)$ (and resp. for the other indexes). Index insertion is shown in Algorithm \ref{alg:indexConstr}. The subroutines in $\mathsf{sans\text{--} serif}$ are described below, by reference to Example \ref{ex:attribute}.

\begin{example}
Let $a$ be an attribute, with name \textit{Address} and extent $[\![$\textit{a}$]\!] = \{ $\footnotesize \texttt{'18 Portland Street,} \texttt{M1 3BE',} \texttt{'41 Oxford Road,} \texttt{M13 9PL'} \texttt{'9 Mirabel Street,} \texttt{M3 1NN'} \normalsize $\}$.
\label{ex:attribute}
\end{example}

\begin{algorithm}[t]
	\caption{Index construction}
	\begin{flushleft}
	\textbf{Input}: Indexes $I_{\mathbb{N}}$, $I_{\mathbb{V}}$, $I_{\mathbb{E}}$, $I_{\mathbb{F}}$, Attribute $a$ \\
 	\textbf{Output}: Updated $I_{\mathbb{N}}$, $I_{\mathbb{V}}$, $I_{\mathbb{E}}$, $I_{\mathbb{F}}$
	\end{flushleft}
	\begin{algorithmic}[1]
		\Function{InsertIntoIndexes}{}
		\State $Q(a) \gets \{\}; T(a) \gets \{\}; R(a) \gets \{\}; \vec{a} \gets \{\};$
		\State $H \gets \mathsf{histogram}.new()$
		\State $Q(a) \gets \mathsf{get\_qgrams}(a)$
		\ForAll{$v \in [\![a]\!]$}
		    \State $H.insert(\mathsf{get\_tokens}(v))$
		    \State $R(a) \gets R(a) \cup \mathsf{get\_regex\_string}(v)$
		\EndFor
		\ForAll{$t \in H.\mathsf{infrequent}()$}
		    \State $T(a) \gets T(a) \cup \{t\}$
		\EndFor
		\ForAll{$t \in H.\mathsf{frequent}()$}
		    \State $\vec{a} \gets \vec{a} \cup \mathsf{get\_embedding}(t)$
		\EndFor
		\State $I_{\mathbb{N}}.insert(\mathsf{MinHash}(Q(a)))$
		\State $I_{\mathbb{V}}.insert(\mathsf{MinHash}(T(a)))$
		\State $I_{\mathbb{E}}.insert(\mathsf{RandomProjections}(\vec{a}))$
		\State $I_{\mathbb{F}}.insert(\mathsf{MinHash}(R(a)))$
		\EndFunction
	\end{algorithmic}
\label{alg:indexConstr}
\end{algorithm}

\begin{itemize}[leftmargin=*,align=left]
\item $\mathsf{get\_qgrams}(a)$: Obtaining the $q$set $Q(a)$ of an attribute $a$ is the straightforward procedure of computing the $q$--grams of its name. We have used $q=4$ as this avoids having too many similar $q$set pair candidates, while benefiting from fine--grained comparisons of attribute names. For Example \ref{ex:attribute}, $\mathsf{get\_qgrams}(a) = \{\mathtt{addr, ddre,dres,ress}\}$. 
\item $\mathsf{frequent}/\mathsf{infrequent}$ tokens: The $t$set $T(a)$ and word--embedding vector $\vec{a}$ of an attribute value are obtained in tandem by construing the extent of $a$ as a set of documents, a value $v$ as a document, each document as a set of parts (split at punctuation characters), and each part as a set of words. With one pass on the extent, we tokenize the values ($\mathsf{get\_tokens}(v)$) and construct a \textsf{histogram} of token occurrences (which we assume to have an associated data structure from which we can retrieve its \textsf{frequent} and \textsf{infrequent} token sets). Then, for each part in the value/document, the procedure (a) adds to $T(a)$, the word $t$ in that part that has the fewest occurrences in the extent, and (b) takes the word in that part that has the most occurrences in the extent, retrieves its word-embedding vector from the WEM\footnote{In machine learning research, many WEMs already exist that vectorize the context in which a word appears in the corpus from which the WEM was built. In this paper, we have used \textit{fastText} \cite{Grave-17} as our WEM.} and adds that vector to $\vec{a}$. For Example \ref{ex:attribute}, $\mathsf{get\_tokens}(a) = \{\mathtt{portland, 3BE, oxford, 9PL, \ldots }\}$. Note that since terms like 'street', 'road', or the area--level tokens in the UK postcode information are frequently occurring they are considered weak signal carriers of value--level similarity, i.e., not part of $T(a)$. However, such terms are indicative of the possible domain--specific types from which the attribute extent is drawn, viz., \textit{Address} in this case. Therefore, they are the terms for which word-embedding vectors are sought, i.e, $\mathsf{get\_embedding}(t)$.
\item $\mathsf{get\_regex\_string}(v)$: The $r$set $R(a)$ of an attribute value builds on the following set of primitive lexical classes defined by regular expressions: $\mathtt{C} = [A-Z][a-z]+$, $\mathtt{U} = [A-Z]+$, $\mathtt{L} = [a-z]+$, $\mathtt{N} = [0-9]+$, $\mathtt{A} = [A-Za-z0-9]+$, $\mathtt{P} = [.,;:/-]+$. $\mathtt{P}$ also includes any other character not caught by previous primitive classes. Given an attribute value, we tokenize it and, for each token $t$, once we find its matching lexical class $l$, we add its denoting symbol to a string that describes the format for the value, and add that string to the set representation $R(a)$. If the same symbol appears consecutively, all occurrences but the first are replaced by '+', e.g., $\{$NC+P+A+$\}$. If an attribute value matches more than one primitive class, we choose the first match, in the order enumerated above.
\end{itemize}

The set representations, obtained as described above, of related attributes are hashed into similar LSH partitions, i.e., rather than indexing full attribute names/values we index set representations, so that signals are both finer--grained and crisper. We can then define $\mathbb{N}$--, $\mathbb{V}$-- and $\mathbb{F}$--relatedness in terms of Jaccard similarity of the corresponding set representations, and $\mathbb{E}$--relatedness in terms of cosine similarity of the corresponding set representations, and efficiently approximate these measures: Jaccard/cosine distance between two set representations is approximated by the bit--level similarity of their MinHash/random--projection values.

\subsection{Attribute Relatedness: The Numeric Case}
\label{sec:numeric-case}

Numeric attributes are a special case in our framework. Of the four types of evidence we take into account, only names and formats provide relatedness evidence when dealing with numbers. This is because numbers cannot be analyzed in terms of tokens as usefully as text can. Hence, token frequency and word--embedding vectors are not useful signals. Moreover, LSH hashing schemes are not available that can be applied to features that we are able to extract from numeric values. So, we do not index numeric values into the respective indexes. We do index them into the name-- and format--related indexes even though, for numbers, formatting  is less indicative of conceptual equivalence. For example, an attribute denoting the age of a person might share many values with an attribute denoting the person's weight or height and it is difficult to think of features that might provide the kind of diversity of viewpoints that we adopt for textual values. In such cases, given two attributes, we ground the decision of relatedness on a distribution similarity measure, the Kolmogorov-Smirnov (KS) statistic \cite{Stats-99}, and use it to decide whether the two corresponding extents, seen as samples, are drawn from the same distribution. 

\begin{algorithm}[t]
	\caption{Computing $\mathbb{D}$-relatedness}
	\begin{flushleft}
	\textbf{Input}: Numeric attributes $a$ in table $T$ and $a'$ in table $S$ \\
	\textbf{Output}: $D_{\mathbb{D}}(a,a')$
	\end{flushleft}
	\begin{algorithmic}[1]
		\Function{Compute$D_{\mathbb{D}}$}{}
			\State $i \gets \mathsf{get\_subject\_attribute}(T)$
		    \State $i' \gets \mathsf{get\_subject\_attribute}(S)$
		    \If{$i' \in I_{\mathbb{*}}.lookup(i)$} \Return $KS([\![a]\!], [\![a']\!])$       
		    \ElsIf{$a' \in I_{\mathbb{N}}.lookup(a)$} \Return $KS([\![a]\!], [\![a']\!])$
		    \ElsIf{$a' \in I_{\mathbb{F}}.lookup(a)$} \Return $KS([\![a]\!], [\![a']\!])$
		    \Else{} \Return 1
		    \EndIf
		\EndFunction
	\end{algorithmic}
\label{alg:D-relatedness}
\end{algorithm}

Algorithm \ref{alg:D-relatedness} describes how we characterize $\mathbb{D}$-relatedness. We use evidence from the $\mathbb{N}$ and $\mathbb{F}$ indexes in a decision on whether we proceed to consider the $\mathbb{D}$--relatedness or not. In addition, in Algorithm \ref{alg:D-relatedness}, we rely on the notion of a \textit{subject attribute} to contextualize the numerical value in terms of the entity of which it is a presumed property. To identify such attributes, we use the supervised learning technique proposed by Venetis \textit{et al.} \cite{Venetis-2011} \footnote{We have built a classification model (invoked in $\mathsf{get\_subject\_attribute}$) and 10-fold cross-validated it on 350 datasets from \textit{data.gov.uk} with manually identified subject attributes. The average accuracy is 89\%.}. Given a dataset, a subject attribute identifies the entities the dataset is about, whereas non-subject attributes describe properties of the identified entity \cite{Venetis-2011, DasSarma-2012}. Intuitively, this approach favours leftmost non-numeric attributes with fewer nulls and many distinct values. As in \cite{DasSarma-2012}, we assume each dataset has only one subject attribute and that this attribute has non-numeric values. 
For example, in Figure~\ref{fig:runex}, the subject attribute of $S_1$ is \textit{Practice Name}, the subject attribute of $S_2$ is \textit{Practice}, the subject attribute of $S_3$ is \textit{GP}, and the subject attribute of $T$ is \textit{Practice}. 
 
We only compute $KS$, our measure of $\mathbb{D}$-relatedness when there is sufficient evidence from indexes we already have that $a$ and $a'$ are related, thereby benefiting from the blocking effect they give rise to. In Algorithm \ref{alg:D-relatedness}, by $I_{*}$ we mean look-ups on all of $I_\mathbb{N}$, $I_\mathbb{V}$, $I_\mathbb{E}$, and $I_\mathbb{F}$, with an existential interpretation, i.e., membership in any one of them. 
 
\subsection{Deciding on Table Relatedness} \label{sec:tabRel}

We have described the types of evidence and corresponding indexes upon which attribute relatedness is defined. We now explain how we use them to return, given a target table and exemplar tuples, the list of its $k$-most related datasets.

Given a target $T$ with attributes $\{a_1, \ldots, a_n\}$, for each $a_i$ we obtain its set representations and use the corresponding hashing schemes to retrieve, from each of the four indexes, each attribute that is related to $a_i$ paired with the corresponding relatedness measure (i.e., its distance to $a_i$). For each related attribute, four distances are returned. If both $a_i$ and the related attribute are numeric, there may be a distribution-based measurement (depending on the guards previously described) computed using the KS statistic, otherwise that measurement is set to 1 (i.e., maximally distant). 

Consider again the example in Figure \ref{fig:runex} where, for each target attribute, we retrieve similar in--lake attributes using the indexes. We group the results by the dataset the attributes originate from. As an example of the structures that are created through this grouping (one for each dataset that has at least one attribute that is related to some target attribute), consider Table~\ref{tab:distEx}. Here, we use hypothetical distance values (the exact ones can be obtained by applying the formulas from Section \ref{sec:attrRel} on the the sets representations of each attribute pair) to exemplify the degree of similarity between attribute pairs. The table contains three rows because, of the five attributes in the target $T$ in Figure \ref{fig:runex}, only three attributes in the $S_2$ dataset are in any degree related to it. Pairs \textit{(T.Practice, $S_2$.Practice}) and \textit{(T.City, $S_2$.City}) have identical attribute names so $D_{\mathbb{N}}$ is 0. For all three pairs in the table, we have $D_{\mathbb{V}}$ and $D_{\mathbb{E}}$ smaller than $1$, which means that there is evidence of their $\mathbb{V}$-- and $\mathbb{E}$--relatedness, and the distribution distance $D_{\mathbb{D}}$ equal to $1$, since all three pairs contain attributes with textual values.

Given the two data sets $T$ and $S2$, in order to compute their relatedness distance, we want to aggregate, column--wise, the distances that appear in the cells of Table~\ref{tab:distEx}, i.e., the distances between their related attributes, to obtain a $5$--dimensional vector that captures the relatedness between the two corresponding datasets. We aggregate using a weighted average of the relatedness distances $\mathcal{D} = \{D_\mathbb{N}, D_\mathbb{V}, D_\mathbb{F}, D_\mathbb{E}, D_\mathbb{D}\}$ to obtain the desired $5$-dimensional vector.

\begin{table}
     \centering
     \small
     \caption{Example Distances for $T$ and $S2$ in Figure \ref{fig:runex}}
     \begin{tabular}{|l|c|c|c|c|c|}
     \hline
     \textit{Pair} & 
     $D_\mathbb{N}$ & $D_\mathbb{V}$ & $D_\mathbb{F}$ & $D_\mathbb{E}$ & $D_\mathbb{D}$ \\  
     \hline
     \hline
     $(T.Practice, S_2.Practice)$ & $0.0$ & $0.9$ & $0.6$ & $0.2$ & $1.0$ \\
     \hline
     $(T.City, S_2.City)$ & $0.0$ & $0.2$ & $0.2$ & $0.3$ & $1.0$ \\
     \hline
     $(T.Postcode, S_2.Postcode)$ & $0.0$ & $0.6$ & $0.1$ & $0.8$ & $1.0$ \\
     \hline
     \end{tabular}
     \normalsize
     \label{tab:distEx}
\end{table}

More formally, let $T$ and $S$ refer to the target and source tables from which Table \ref{tab:distEx} is constructed. We use Equation~\ref{eq:wavg} on each column of Table \ref{tab:distEx} to aggregate its values:

\begin{equation} \label{eq:wavg}
\small
    D_{t}(T,S) = \frac{\sum_{i=1}^{m} w_t^iD_t^i}{\sum_{i=1}^{m} w_t^i}
\end{equation}
with $m$, the number of attributes in $T$ that are related to some attribute in $S$, and $t \in \{\mathbb{N}, \mathbb{V}, \mathbb{F}, \mathbb{E}, \mathbb{D}\}$.

We must define the weights to use in Equation \ref{eq:wavg}. Recall that, for each distance type $t$, by performing a look--up on the corresponding index for a target attribute $a$, we retrieve every attribute of datasets in the lake that is related to $a$, paired with the corresponding relatedness measure, i.e., its distance to $a$ computed as described in Section \ref{sec:indexConstr}. In other words, for each target attribute $a$, we can compute a distribution of relatedness measurements of type $t$, i.e., the set of all distances of type $t$ between $a$ and every attribute in the lake that is related to $a$. We denote such a set as $\mathcal{R}_t$. Given a distance value $D^i_t$ between two attributes, e.g., a cell value from Table \ref{tab:distEx}, its associated weight $w_t^i$ is given by the complementary cumulative distribution function evaluated at $D^i_t$:

\begin{equation} \label{eq:weights}
\small
    \begin{aligned}
    w_t^i = 1 - P(d \leq D^i_t), \forall d \in \mathcal{R}_t
    \end{aligned}
\end{equation}

Intuitively, each weight $w_t^i$ represents the probability that the observed distance $D^i_t$ is the smallest in $\mathcal{R}_t$. This allows the weights to compensate for the presence of a potentially high number of weakly related attributes to a target attribute.

As an example, consider again the pair of datasets $(T, S_2)$ from Figure \ref{ex:intro}, with their aligned attributes shown in Table \ref{tab:distEx}. For each $t \in \{\mathbb{N}, \mathbb{V}, \mathbb{F}, \mathbb{E}, \mathbb{D}\}$, we use the distribution, $\mathcal{R}_t$, of all computed distances of type $t$ between the target attribute and all other $t$--related attributes in the lake, to decide how important a given distance $D_t^i$ is in Equation \ref{eq:wavg}. For instance, if $S_2.Postcode$ is among the most $\mathbb{V}$--related attributes to $T.Postcode$ in the entire data lake, $D_{\mathbb{V}}^3$ (i.e., the third value on $D_{\mathbb{V}}$ column of Table \ref{tab:distEx}) will have a high weight in Equation \ref{eq:wavg} denoting a strong relatedness signal, relative to all other attributes $\mathbb{V}$--related to $T.Postcode$. Conversely, if $S_2.Postcode$ is among the least $\mathbb{E}$--related attributes to $T.Postcode$ in the entire data lake, $D_{\mathbb{E}}^3$
will have a low weight in Equation \ref{eq:wavg} denoting a weak relatedness signal, relative to all other $\mathbb{E}$--related attributes to $T.Postcode$.

Equation~\ref{eq:wavg} is applied on each column of a table like Table \ref{tab:distEx}, and this results in a $5$-dimensional vector, $\vec{dv}_{(T,S)} = [D_\mathbb{N}(T,S), D_\mathbb{V}(T,S), D_\mathbb{F}(T,S), D_\mathbb{E}(T,S), D_\mathbb{D}(T,S)]$. In order to derive a scalar value from $\vec{dv}_{(T,S)}$ that can stand as a measurement of the relatedness between $T$ and $S$, we consider $S$ to be a point in a $5$--dimensional Euclidean space, where each distance measure represents a different dimension. In this space, the coordinates of $T$ are $[0, 0, 0, 0, 0]$. This allows us to compute a combined distance of $S$ from $T$ using the weighted $l^2$--norm of $\vec{dv_{(T,S)}}$ (i.e., the weighted euclidean distance):

\begin{equation} \label{eq:euclid}
\small
     D(T,S) = \sqrt{\frac{\sum_{t=1}^5 (w_t \times \vec{dv}_{(T,S)}[t])^2}{\sum_{t=1}^5 w_t}}
 \end{equation}

Again, we must define the weights to use in Equation~\ref{eq:euclid}. Note that here the weights represent a proposal as to the relative importance of each evidence type $t$, i.e., each type of relatedness measure. We started by construing relatedness discovery as a binary classification problem. Then:

\begin{enumerate}[leftmargin=*,align=left]
\item We used the benchmark provided in \cite{Nargesian-2018}, which comes with the ground truth about relatedness, to create a training set by choosing related and unrelated pairs of the form $(T, S)$ (i.e., positive and negative examples, resp.) from the benchmark ground truth. In the training set, if $S$ is related to $T$, then we label the pair as \textit{related} (i.e., 1), otherwise we label it as \textit{unrelated} (i.e., 0). Each such pair has a feature vector of five associated elements, i.e., the five distance measures obtained through Equation \ref{eq:wavg}. 
\item We built a logistic regression classifier using the training set, relying on \textit{coordinate descent} \cite{Hsieh-08} to optimize the coefficient of each feature. We tested the resulting model against a test set, created similarly to the training set, using data from a ground truth of a manually created real--world benchmark, and obtained an accuracy of approx. $89\%$. Details about the test benchmark are given in Section \ref{sec.eval}.
\item We used the coefficients of the resulting model as the respective weights in Equation \ref{eq:euclid}. 
\end{enumerate}

The intuition is that the classifier coefficients will minimize the distance between highly--related datasets and maximize it between unrelated datasets. 

Given a target table $T$ to be populated, and a data repository $\mathcal{S} = \{S_1, S_2, ... , S_n\}$, the \textit{dataset discovery} problem is the problem of finding the $k$-most related datasets to $T$ in $\mathcal{S}$, where dataset relatedness is measured using Equation \ref{eq:euclid}. 

\section{Extending relatedness through join paths} \label{sec:joins}

The techniques described so far construe relatedness discovery as finding datasets in the lake with attributes that are aligned (by which we mean `related by any of the evidence types') to as many attributes in the target as possible. In this section, we show how some of the indexes we build for characterizing similarity can be used to discover join opportunities between the $k$--most related tables to a target and non--top--$k$ tables. Thus, tables with weaker relatedness signal are included in the solution if, through joins, they contribute to covering more attributes in the target.

Given a target $T$, let $\mathcal{S}=\{S_1, \ldots S_n\}$ be the set of all datasets from a data lake, and $\mathcal{S}^k, k\leq n$, the $k$-most related datasets to $T$. In this section, we describe how we identify datasets in $\mathcal{S} - \mathcal{S}^k$ that, through joins with datasets in $\mathcal{S}^k$, contribute to populating $T$.

\begin{algorithm}[t]
	\caption{Join paths discovery}
	\begin{algorithmic}[1]
		\Function{FindJoinPath}{$start$, $path$}
		    \State $path.append(start)$
		    \ForAll{$N_i \in G_{\mathcal{S}}.neighbours(start)$}
		    \If{$N_i \notin \mathcal{S}^k \ \&\& \ N_i \notin path \ \&\& \ N_i \in I_{\mathbb{*}}.lookup(T)$}
		        \State $new\_path \gets \textproc{FindJoinPath}(N_i, path)$
		        \State $\mathcal{J} \gets \mathcal{J} \cup \{new\_path\}$
		    \EndIf
		    \EndFor
		    \State \Return $path$
		\EndFunction
	\end{algorithmic} 
    \label{alg:joinPath}
\end{algorithm}

We focus on joins based on postulated (possibly partial) inclusion dependencies. Although these can be computed using data profiling techniques \cite{Papenbrock-15}, this is not practical given the size of the data lakes we are focusing on, i.e., the size of the all-against-all attributes search space. As such, we consider two datasets $S$ and $S'$ to be \textit{joinable} if they are \textit{SA--joinable}  (\textit{SA} for subject--attribute: described in Section \ref{sec:numeric-case}) and we consider $S$ and $S'$ to be \textit{SA--joinable} if (i) there is $I_{\mathbb{V}}$-based evidence that the $t$sets $T(a)$ and $T(a')$, where $a$ and $a'$ are attributes of $S$ and $S'$, resp., overlap, and (2) at least one of $a$ or $a'$ is a subject attribute. Thus, we rely on $I_{\mathbb{V}}$ to identify inclusion dependencies and, instead of the notion of candidate key, we use subject attributes.

To determine whether two $t$sets overlap, we define the \textit{overlap coefficient} between two $t$sets as follows $ov(T(a), T(a')) = \frac{|T(a) \cap T(a')|}{\min (|T(a)|,|T(a')|)}$. Let $\tau$ be the similarity threshold parameter configured for LSH, i.e., if $a$ and $a'$ are $\mathbb{V}$--related, given the properties of LHS under MinHash, then they are Jaccard--similar with a similarity between their respective $t$sets $\ge \tau$. Then, $\frac{\tau \times (|T(a)| + |T(a')|)}{(1 + \tau) \times \min (|T(a)|,|T(a')|)} \le ov(T(a), T(a')) \le 1$, by the set-theoretic inclusion-exclusion principle.

We construe the discovery of join paths as a \textit{graph traversal} problem, and, in order to identify \textit{SA--join} paths among the elements of $\mathcal{S}$, we define an \textit{SA-join graph}, $G_{\mathcal{S}} = (\mathcal{S}, \mathcal{I})$ over the entire data lake, where $\mathcal{S}$ is the node set and $\mathcal{I}$ the edge set defined using the two \textit{SA--joinability} conditions from above: each edge from $\mathcal{I}$ connects two \textit{SA--joinable} nodes $S_i$ and $S_j$.

Given an \textit{SA--join} graph $G_{\mathcal{S}}$, and a set $\mathcal{S}^k$ of $k$--most related datasets to a target $T$, we find the set of \textit{SA-join} paths from each $S_i \in \mathcal{S}^k$ to all other vertices in $G_{\mathcal{S}}$ (or in a connected component of $G_{\mathcal{S}}$ that contains $S_i$) that are not in $\mathcal{S}^k$, using Algorithm \ref{alg:joinPath}. Specifically, the function traverses $G_{\mathcal{S}}$ depth--first, starting from $S_i$ and adds join paths to a globally accessible set $\mathcal{J}$ whenever (i) all path nodes, apart from the starting node $S_i$, are not in $\mathcal{S}^k$, (ii) the path is not cyclic, and (iii) there is evidence from at least one index that every node in the path is related to the target.

Algorithm \ref{alg:joinPath} is called for each $S_i \in \mathcal{S}^k$ and returns a set of \textit{SA--join} paths of variable lengths, each of which starts from $S_i$. Each dataset in such a join path has the potential to improve target population, either through the addition of new instance values to an already covered target attribute, or by populating previously uncovered target attributes. Our experimental results show that, by taking join opportunities into account, both the achievable ratio of covered target attributes and the precision of attributes that are considered for populating the target are improved.

\section{Evaluation} \label{sec.eval}

We firstly evaluate the effectiveness of each relatedness evidence type and compare them against the aggregated approach which considers all of them. We then compare the effectiveness and the efficiency of $D^3L$ with that of the techniques proposed in \cite{Nargesian-2018} (referred to as $TUS$ for Table Union Search) and in \cite{Fernandez-18} (referred to as $Aurum$). Finally, we evaluate the impact on target coverage and precision when, in addition to the top--$k$, we also consider datasets that are joinable with tables in the top--$k$. We use the following repositories in the experiments:
\begin{itemize}[leftmargin=*]
    \setlength\itemsep{.1em}
    \item \textit{Synthetic} ($\sim$1.1GB): $\sim$5,000 tables (used in $TUS$ \cite{Nargesian-2018}) synthetically derived from 32 base tables containing Canadian open government data using random projections and selections on the base tables. We use this repository to measure comparative effectiveness in terms of precision and recall. The average answer size is 260 (i.e., the average number of related tables over 100 randomly picked targets). This dataset is available from:  \url{github.com/RJMillerLab/table-union-search-benchmark.git}.
    \item \textit{Smaller Real} ($\sim$600MB): $\sim$700 tables from real world UK open government data, with information on domains such as business, health, transportation, public service, etc. Again, we use this repository to measure comparative effectiveness. The average answer size is 110.
    \item \textit{Larger Real} ($\sim$12GB): $\sim$43,000 tables with real world information from different UK National Health Service organizations (\url{webarchive.nationalarchives.gov.uk/search/}).  We only use this repository to measure comparative efficiency\footnote{The scripts used to download the two real world data sets are available from: \url{github.com/alex-bogatu/DataSpiders.git}}.

\end{itemize}   

For \textit{Synthetic}, the ground truth resulted from recording for every table, through the derivation procedure, which other tables are related to it. For \textit{Smaller Real} a human has manually recorded, for every table in the lake, which other tables are related to it, as defined in Definition \ref{def:relatedness}. In both ground truth instances each table $T$ in the repository is listed with all its attributes along with every table $T'$, along with its own attributes that are related to some attribute in $T$. As per Definition \ref{def:relatedness}, two attributes are considered related in the ground truth if both contain values drawn from the same domain.

\begin{figure*}[t]
\centering
\begin{subfigure}{.3\textwidth}
    \centering
    \includegraphics[width=\linewidth, trim=2 2 2 2,clip]{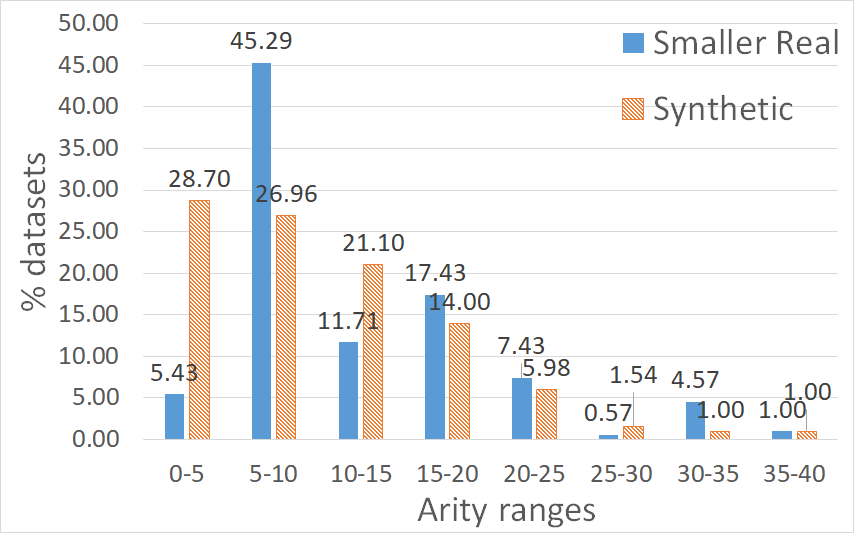}
    \caption{Arity}
    \label{fig:arity}
\end{subfigure}
\begin{subfigure}{.3\textwidth}
    \centering
    \includegraphics[width=\linewidth, trim=2 2 2 2,clip]{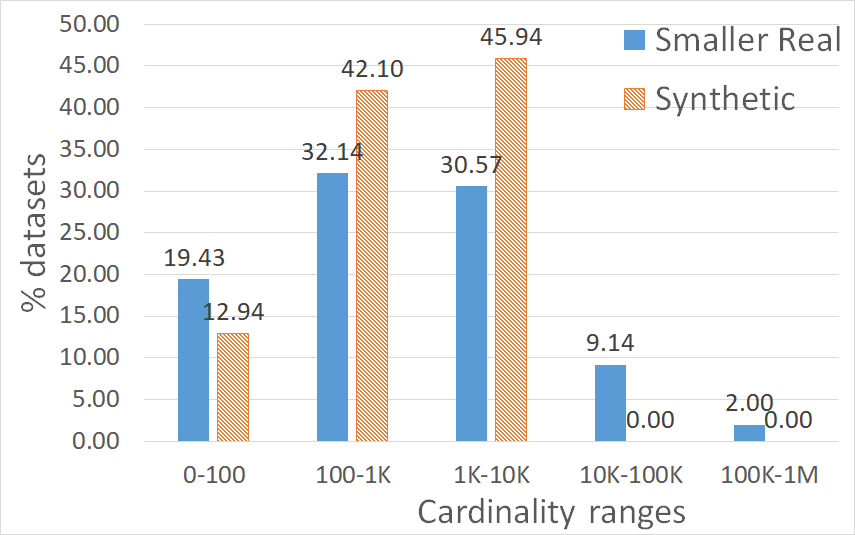}
    \caption{Cardinality}
    \label{fig:cardinality}
\end{subfigure}
\begin{subfigure}{.3\textwidth}
    \centering
    \includegraphics[width=\linewidth, trim=2 2 2 2,clip]{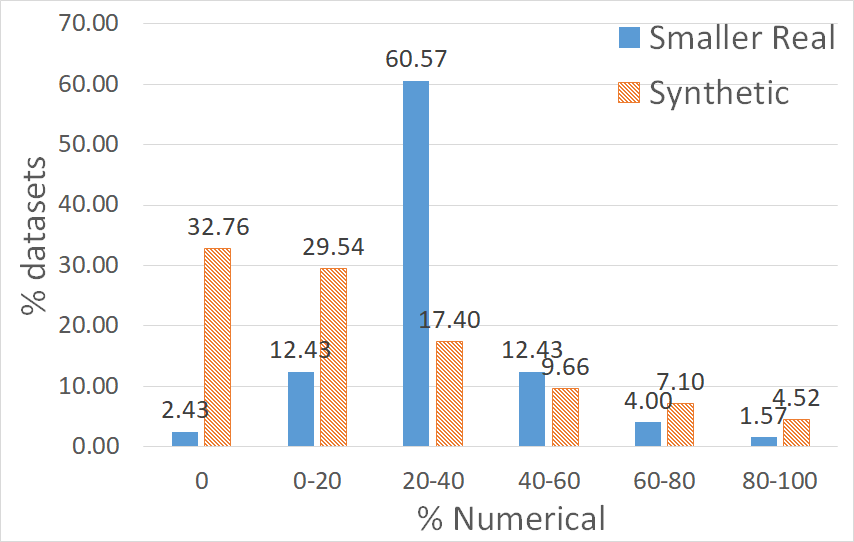}
    \caption{Data types}
    \label{fig:numerical}
\end{subfigure}
\caption{\textit{Synthetic} and \textit{Smaller Real} statistics}
\label{fig:stats}
\end{figure*}

Figure \ref{fig:stats} describes the arity, the cardinality and the percentage of numerical attributes of the two repositories used in measuring effectiveness. Arity can have a significant impact on the top-$k$ ranking, i.e., sources with many similar attributes tend to be ranked higher by our weighted scheme, and on target coverage, i.e., the number of attributes related to some target attribute. Cardinality influences the accuracy of similarity estimation and of join path discovery, i.e., a high overlap between instance values determines a high probability of collisions between MinHash hashes.
Lastly, numerical attributes are an important special case, as discussed in Section \ref{sec:numeric-case}.

\subsection{Baselines and reported measures} \label{subsec.baseline}

$TUS$ \cite{Nargesian-2018} proposes a unionability measuring framework that builds on top of three types of evidence extracted exclusively from instance--values, aiming to inform decisions on unionability between datasets from different viewpoints. $TUS$ uses similar indexing and querying models to $D^3L$ and, therefore, it is a good candidate for a comparative analysis against $D^3L$ w.r.t. both effectiveness and efficiency.

$Aurum$ \cite{Fernandez-18} uses both schema-- and instance--level information to identify different relationships between attributes of a data lake. A two--step process profiles and indexes the data, creating a graph structure that can be used for key--word search, unionability, or joinability discovery. This makes $Aurum$ a good candidate for a comparative analysis against $D^3L$  w.r.t. indexing time, effectiveness\footnote{For $Aurum$, we use the \textit{certainty} ranking strategy described in \cite{Fernandez-18}, i.e., when attributes are related by more than one evidence type, similarly to $TUS$, the maximum similarity score gives the value used in ranking the results.}, and the added value of join paths. Conversely, $Aurum$ employs a different querying model, treating queries as graph traversal problems, rather than LSH index lookups. This means that the discovery process in $Aurum$ is not influenced by the same parameters, e.g., $k$, as in $D^3L$. This makes an efficiency comparison between $Aurum$ and $D^3L$ w.r.t. search time infeasible.

Given a target $T$, we report the \textit{precision} and \textit{recall} of the top--$k$ datasets related to $T$, because, in this experiment, we are not interested in the, potentially many, data lake members weakly related to $T$, but only in the top--$k$.

For the purposes of computing precision and recall, we define a true positive, $TP$: a table from repository $R$ that is in the top-$k$ tables returned and is related to the target in the ground truth for $R$; a false positive, $FP$: a table from repository $R$ that is in the top-$k$ tables returned and is not related to the target in the ground truth for $R$; and a false negative, $FN$: a table from repository $R$ that is related to the target in the ground truth for $R$ but is not a member of the top-$k$ tables returned. As usual, precision $p = TP/(TP+FP)$ and recall $r = TP/(TP+FN)$. In assessing the result (i.e., the top-$k$ tables returned), we count the occurrence of a table in the answer as a true positive if, as per the corresponding ground truth, at least one, but not necessarily all, attributes of a table in the solution  is related to the target. 

In our interpretation of true positives, we consider that failing to identify one related attribute should not be considered a sufficient condition for concluding that the table it belongs to is unrelated to the target: every attribute that can contribute to populating the target does indeed so contribute. We present more insight on the coverage of the target in the experiments pertaining to relatedness as joinability, i.e., Experiments 8--11.

In the results\footnote{Each solution, \textit{viz.} $D^3L$, $TUS$, and $Aurum$, is implemented using LSH Forest\cite{Bawa-05} configured with a threshold of $0.7$ and a MinHash size of $256$. All experiments have been run on Ubuntu 16.04.1 LTS, on a 3.40 GHz Intel Core i7-6700 CPU and 16 GB RAM machine.} below, each point is the average computed by running $D^3L$ (and $TUS$/$Aurum$, where pertinent) over 100 randomly selected targets from the respective repository.

\subsection{Individual effectiveness}

We first evaluate the effectiveness of data discovery conducted using each $D^3L$ evidence type individually. We only discuss here the results obtained for the \textit{Smaller Real} repository; running the experiment on the \textit{Synthetic} repository returned similar behaviour in terms of precision and recall.

\begin{figure}[t]
\centering
\begin{subfigure}{.22\textwidth}
    \centering
    \includegraphics[width=\linewidth, trim=4 4 3 4,clip]{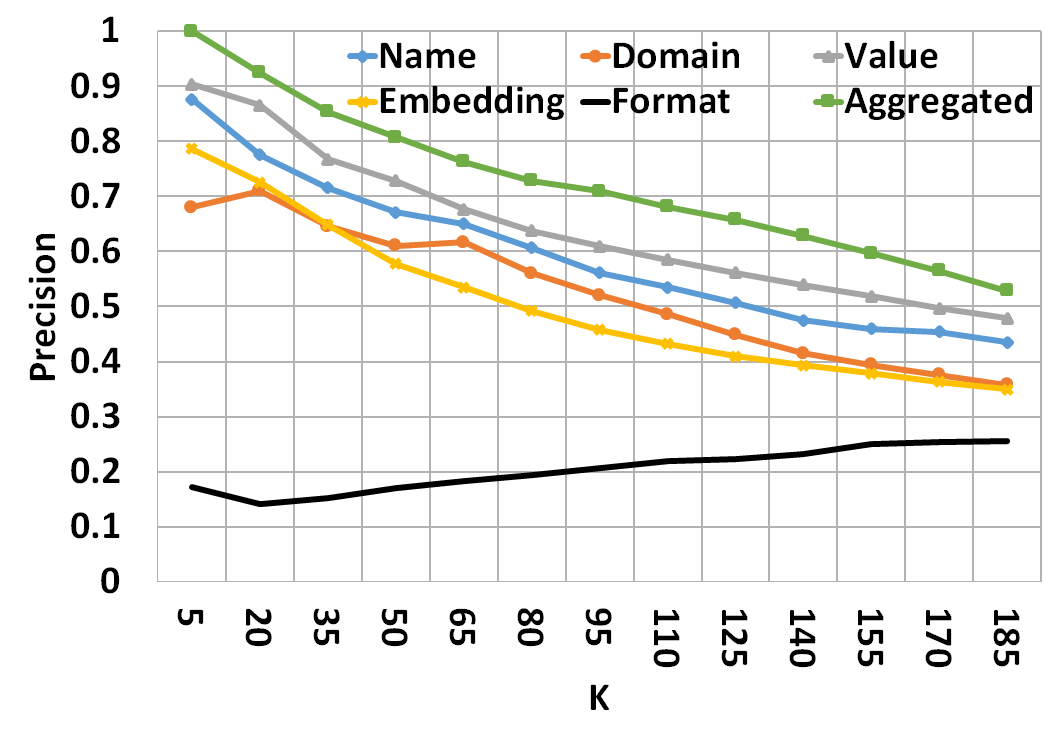}
    \caption{Precision}
    \label{fig:indivP}
\end{subfigure}
\begin{subfigure}{.25\textwidth}
    \centering
    \includegraphics[width=\linewidth, trim=4 4 3 4,clip]{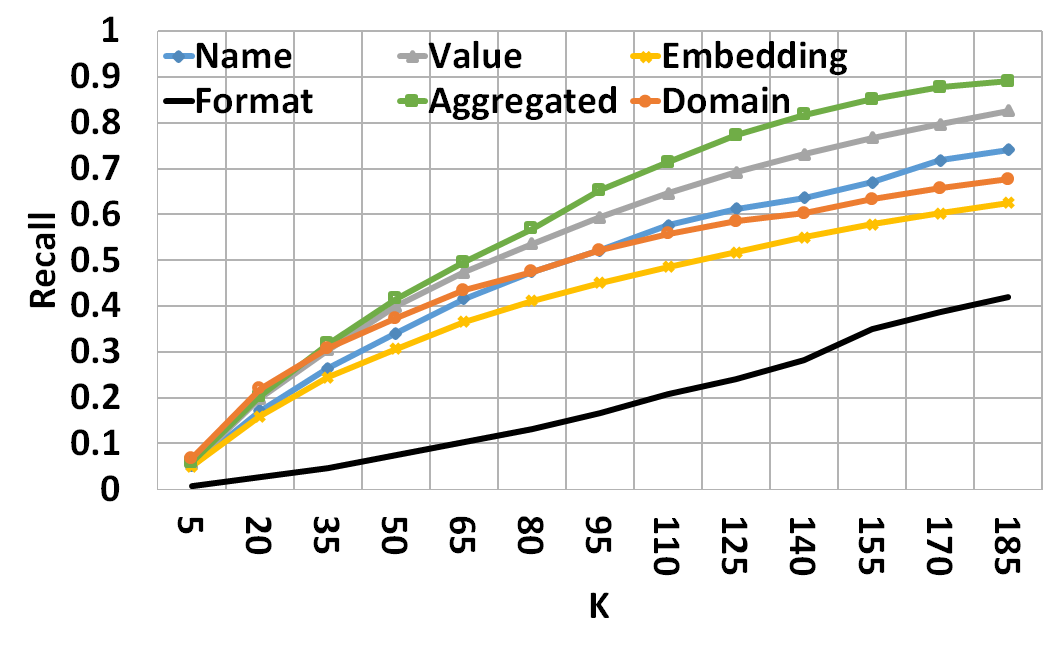}
    \caption{Recall}
    \label{fig:indivR}
\end{subfigure}
\caption{Individual Precision and Recall on \textit{Smaller Real} }
\label{fig:indivPR}
\end{figure}

\smallskip \noindent \textbf{Experiment \stepcounter{experiment}\theexperiment\xspace: Precision and recall (on \textit{Smaller Real}) for each type of evidence, as answer size grows}.
The purpose of this experiment is to evaluate the effectiveness of individual evidence types against what is achievable when using the combined approach. In Figure \ref{fig:indivPR}, the low precision (e.g., $[0.10, 0.30]$) and recall (e.g., $[0.03, 0.43]$) achieved using \textit{format} suggests that the format representation in itself is not sufficiently discriminating, e.g., there may be many single--word or number attributes that represent different entities. The remaining evidence types yield higher precision: at the average answer size, $k=110$, all four evidence types achieve between $0.43$ (\textit{embeddings}) and $0.60$ (\textit{values}) precision, and between $0.49$ (\textit{embeddings}) and $0.70$ (\textit{values}) recall. 

Aggregating all five measures, using the aggregation framework described in Section \ref{sec:tabRel}, results in a nearly constant increase in both measures, compared with the best individual evidence type: \textit{values}. For instance, at $k=110$, the $60\%$ precision achieved when using \textit{value}--based similarity increases to almost $70\%$ when considering all evidence types. Similarly, recall increases from $65\%$ when using values to more than $70\%$ when combining all measurements. Overall, there is a $29\%$ increase in the percentage of correct values returned at $k=110$, which explains the increase in both precision and recall when all relatedness evidence types are considered. 

Performing the same experiment for non--numerical attributes only, i.e., $D_{\mathbb{D}} = 1$, resulted in an average decrease in the aggregated precision and recall of less than $3.5\%$ each. This suggests that, for this benchmark, most of the discoverable relatedness relationships between numerical attributes are already identified by other types of evidence, e.g., $\mathbb{N}$, $\mathbb{F}$.

\subsection{Comparative Effectiveness} \label{sec:compEfec}

\begin{figure}[t]
\centering
\begin{subfigure}{.23\textwidth}
    \centering
    \includegraphics[width=\linewidth, trim=4 4 4 4,clip]{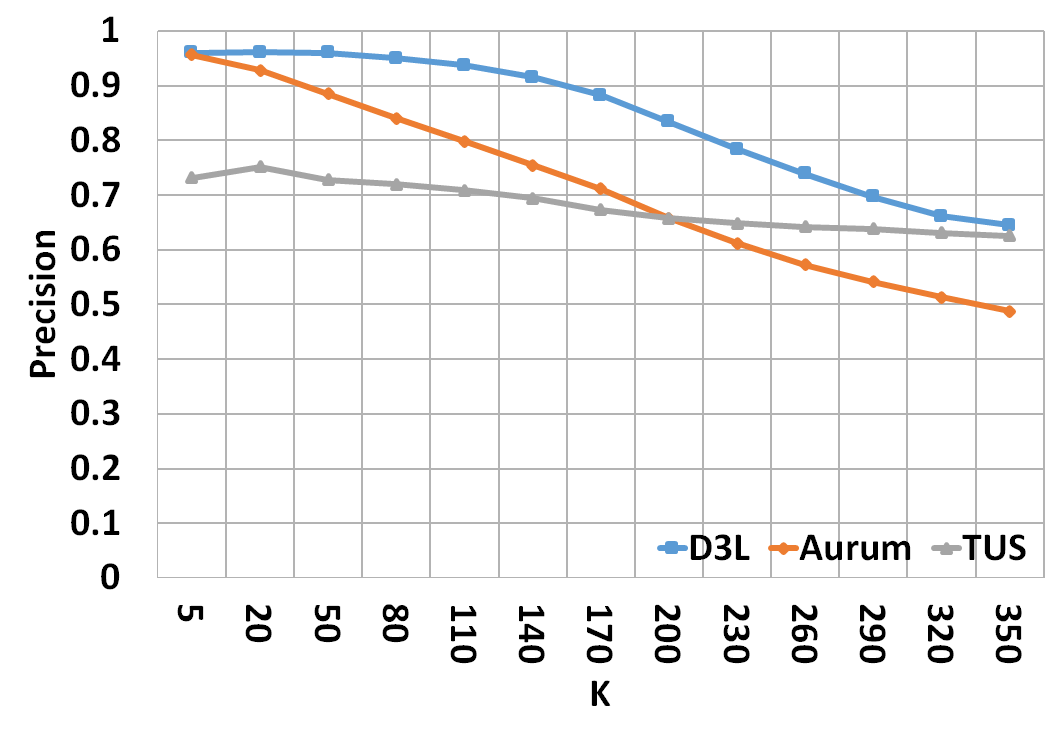}
    \caption{Precision}
    \label{fig:synth_p}
\end{subfigure}
\begin{subfigure}{.23\textwidth}
    \centering
    \includegraphics[width=\linewidth, trim=4 4 4 4,clip]{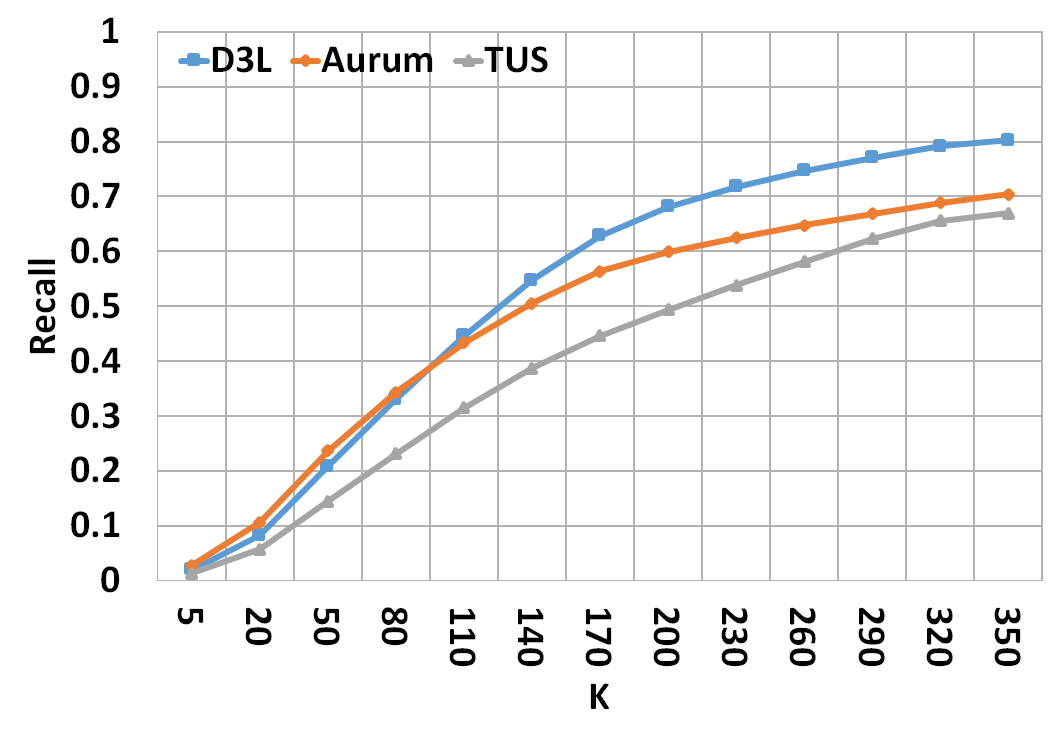}
    \caption{Recall}
    \label{fig:synth_r}
\end{subfigure}
\caption{Precision and Recall on \textit{Synthetic}}
\label{fig:synthPR}
\end{figure}

In this experiment we report the precision and recall of $D^3L$, $TUS$ and $Aurum$, at $k$ (i.e., computed over the top-$k$ tables returned) on the \textit{Synthetic} and \textit{Smaller Real} repositories.

\smallskip \noindent \textbf{Experiment \stepcounter{experiment}\theexperiment\xspace: Precision and recall (on \textit{Synthetic}) as answer size grows}. 
Figure \ref{fig:synthPR} shows $D^3L$ to be highly precise for $k \in [5,140]$ and to linearly decrease in the second part of the interval (down to $0.65$ when $k=350$). This suggests that most of the closely related datasets are at the top of the ranking. Similarly, $Aurum$ is comparatively precise for $k \in [5,50]$ but degrades linearly for the rest of the interval (down to $0.49$ when $k=350$). $TUS$ precision suggests that between $20\%$ and $30\%$ of the retrieved results are false positives consistently ranked higher than truly related tables. 

Overall, $D^3L$ performs better that the baselines because the finer--grained features are more diagnostic of similarity and the aggregation framework allows each evidence to contribute to the ranking, therefore reducing the impact of highly--scored false positives, i.e., a strong score in one dimension is balanced with a, potentially, lower score in another. By contrast, both baselines employ a \textit{max--score} aggregation that only considers the highest similarity score. In case of $TUS$, the transformation of similarity scores into probabilities determines a further dispersion of true positives across the entire set of results.

Recall rises fast for $k \in [5,140]$ for all approaches and levels out beyond the average answer size. As $k$ increases, $D^3L$ is able to identify up to $20\%$ more relevant tables compared to $TUS$, and up to $10\%$ more relevant tables compared to $Aurum$. This is because $D^3L$ employs a multi--evidence relatedness discovery that guards against too many misses. We found that both $TUS$ and $Aurum$ tend to miss relevant attributes that do not share values with some target attribute. This is because $TUS$ relies exclusively on instance values evidence, and $Aurum$'s name and TF/IDF--based evidence proves less dependable than content--based evidence.

\begin{figure}[t]
\centering
\begin{subfigure}{.23\textwidth}
    \centering
    \includegraphics[width=\linewidth, trim=4 4 4 4,clip]{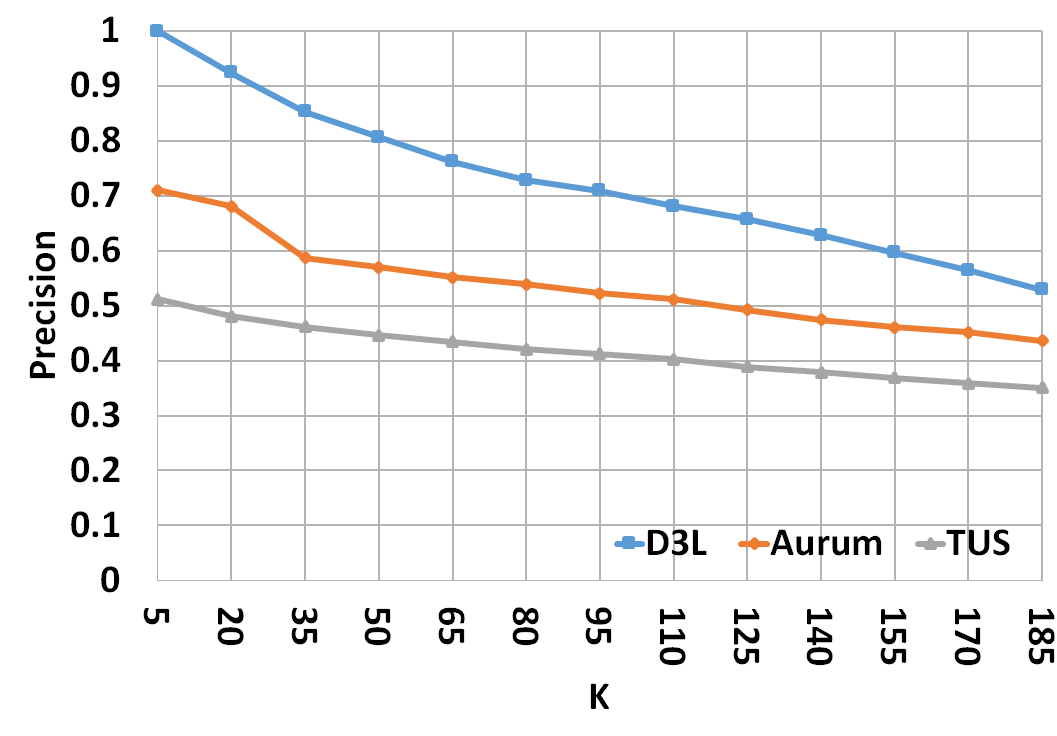}
    \caption{Precision}
    \label{fig:real_p}
\end{subfigure}
\begin{subfigure}{.23\textwidth}
    \centering
    \includegraphics[width=\linewidth, trim=4 4 4 4,clip]{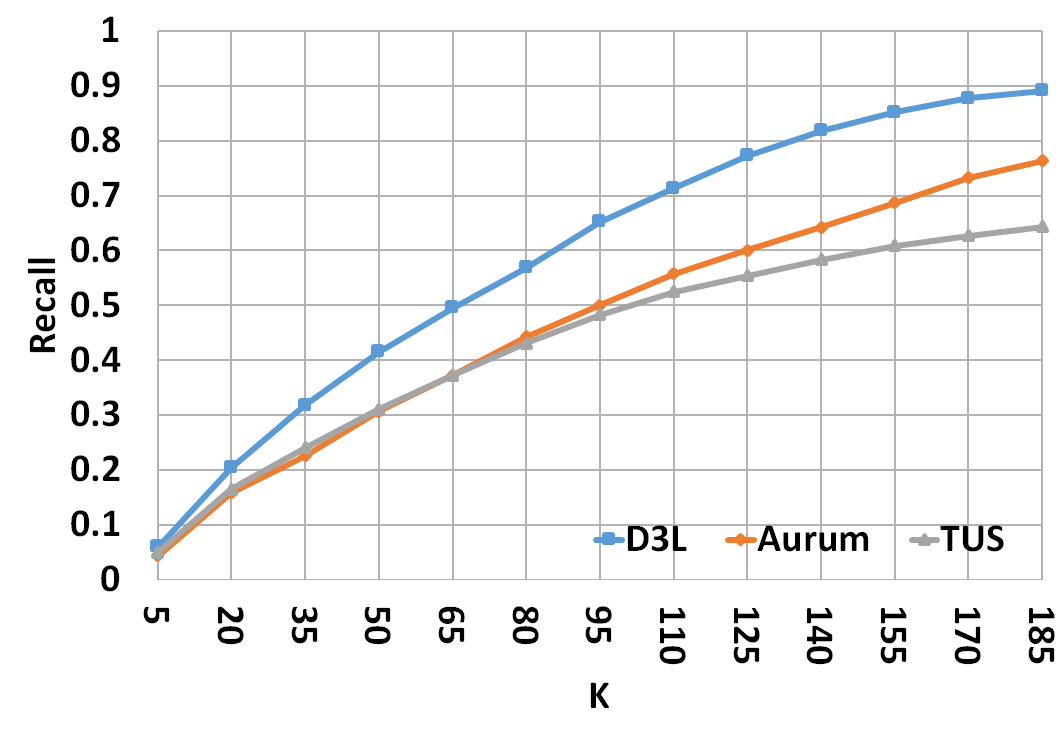}
    \caption{Recall}
    \label{fig:real_r}
\end{subfigure}
\caption{Precision and Recall on \textit{Smaller Real}}
\label{fig:realPR}
\end{figure}

\smallskip \noindent \textbf{Experiment \stepcounter{experiment}\theexperiment\xspace: Precision and recall (on \textit{Smaller Real}) as answer size  grows.}
Figure \ref{fig:realPR} shows that $D^3L$ correctly identifies highly related datasets, e.g., $k \in [5,110]$, resulting in precision between $0.2$ and $0.4$ higher compared to $TUS$, and between $0.05$ and $0.3$ higher compared to $Aurum$. This is because the value--based similarity evidence used by $TUS$ and $Aurum$ expect equality between the instance values of similar attributes, which is not a characteristic of \textit{Smaller Real}. As with the \textit{Synthetic} benchmark, the aggregation framework of $D^3L$ contributes to the improved precision as well.


Regarding recall, at the average answer size ($k=110$), $D^3L$ identifies more than $70\%$ of the related datasets, while both $TUS$ and $Aurum$ identify around $55\%$. The performance gap is wider between $D^3L$ and the two baselines for \textit{Smaller Real} than for \textit{Synthetic} across the entire range of $k$ values. This is because $D^3L$ employs a more lenient approach w.r.t. format representation of values when indexing and comparing attributes. In contrast, $TUS$ and $Aurum$ are more dependent on consistent, clean values than $D^3L$. High levels of consistency and cleanliness are features of the synthetically generated tables but are less prevalent in the real tables.


\subsection{Comparative Efficiency}

We report performance data for $D^3L$, $TUS$, and $Aurm$, where pertinent. We report the time it takes to create the indexes and the time it takes to compute the top-$k$ solution. Note that the implementation of $TUS$ in \cite{Nargesian-2018} is not publicly available so we have implemented it ourselves using information from the paper. For $Aurum$, we have used the implementation from \textit{\url{github.com/mitdbg/aurum-datadiscovery}}

\smallskip \noindent \textbf{Experiment \stepcounter{experiment}\theexperiment\xspace: Time to create the indexes as the data lake size grows.}  
For this experiment, we use the \textit{Larger Real} repository to enable the evaluation on a wider range of data lake sizes. We took five random samples from it at sizes starting from $2.5K$ tables and $20K$ attributes, growing the repository by $2.5K$ and $20K$, resp., at each step.

\begin{figure*}[t]
\centering
\begin{subfigure}{.35\textwidth}
    \centering
    \includegraphics[width=\linewidth]{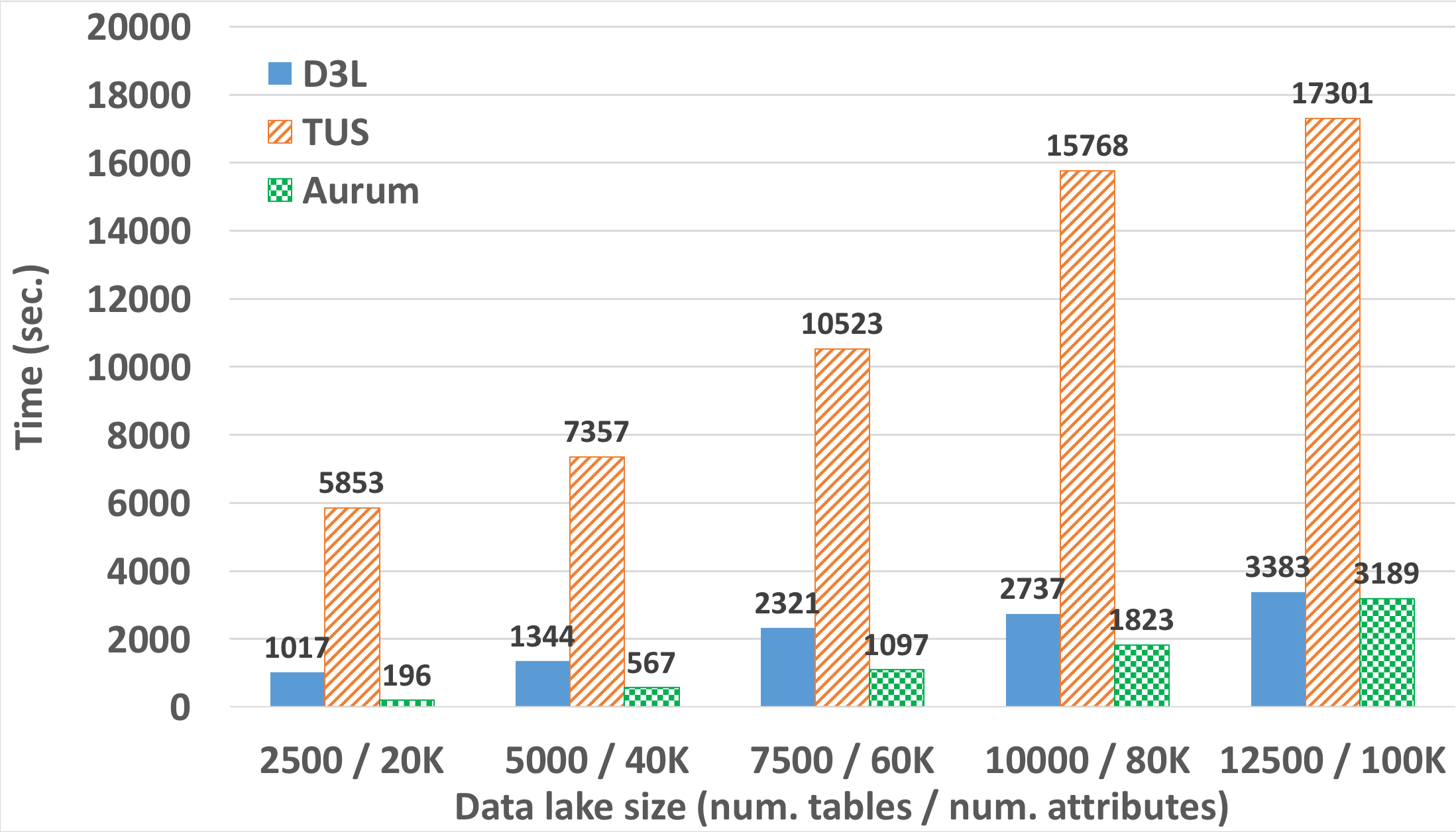}
    \caption{Indexing time}
    \label{fig:indexing}
\end{subfigure}
\begin{subfigure}{.3\textwidth}
    \centering
    \includegraphics[width=\linewidth, trim=2 2 30 2,clip]{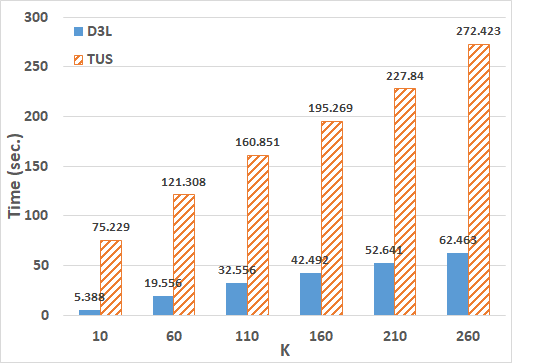}
    \caption{\textit{Synthetic} search time}
    \label{fig:synth_search}
\end{subfigure}
\begin{subfigure}{.3\textwidth}
    \centering
    \includegraphics[width=\linewidth, trim=2 2 30 2,clip]{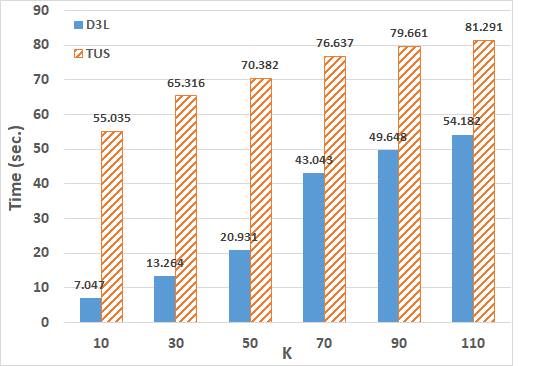}
    \caption{\textit{Smaller real} search time}
    \label{fig:real_search}
\end{subfigure}
\caption{Indexing and searching performance}
\label{fig:efficiency}
\end{figure*}

The results are shown in Figure \ref{fig:indexing}. For each system, the reported values include the times required for pre--processing the data and for creating all data structures later used in performing dataset discovery.

Compared to $TUS$, $D^3L$ performed up to $4$x better on small and medium sized lakes, e.g., $7.5K$ tables, and up to $6$x better on larger ones, e.g., $12.5K$. $Aurum$ performs up to $5$x better that $D^3L$ for small data lakes, e.g., $2.5K$ tables, and comparable with $D^3L$ for larger lakes, e.g., $12.5K$. The dominant task in both $D^3L$ and $TUS$ is data pre--processing, e.g., generating summary representations for each attribute, while in $Aurum$ the dominant task is the creation of the graph structure used to perform discovery. The common tasks of generating \textit{MinHash}/random--projection signatures and creating LSH indexes have been found to take comparable amounts of time in all three systems. The main reason for the poorer performance of $TUS$ seems to lie in its approach to semantic evidence, for which, in \cite{Nargesian-2018}, YAGO \cite{Suchanek-07} is used. Having to map each token of each instance value into a YAGO knowledge base significantly slows down index construction and, as the effectiveness results have shown, for perhaps insufficient return on investment.



\smallskip \noindent \textbf{Experiment \stepcounter{experiment}\theexperiment\xspace (on \textit{Synthetic}): Effect on search time as answer size grows.} In the next two experiments we report the effect of the answer size on the time needed to compute the answer. The requested size of the answer is the parameter that most significantly affects the search time for $D^3L$ and $TUS$. Conversely, the $Aurum$ query model is not impacted by the size of the result: even when using LSH Forest, the indexes are queried only once, when the graph structure is created. In the case of $D^3L$ and $TUS$, every query is an index lookup task parametrized with a value for $k$ (the answer size).

Figure \ref{fig:synth_search} shows the results for \textit{Synthetic}. $D^3L$ performs much better than $TUS$ because the reliance on YAGO of the latter to provide semantic information proves to be a performance leakage point: recall that, at search time, the same process of mapping each instance value to YAGO is applied on the target. Moreover, in $TUS$, the index is only a blocking mechanism, i.e., there remains a significant amount of computation to be done before the unionability measurements are obtained. In contrast, $D^3L$ does not use knowledge--base mapping and its distance--based approach means that search returns plug directly into relatedness measurements. 

Although not directly comparable, we also report the average search time of $Aurum$ obtained for 100 queries on \textit{Synthetic}, with a graph structure that accommodates a result size of at least $260$ datasets: $22.42$ seconds.

\smallskip \noindent \textbf{Experiment \stepcounter{experiment}\theexperiment\xspace (on \textit{Smaller Real}): Effect on search time as answer size grows.} 
In this experiment we use the \textit{Smaller Real} for which we vary the answer size from $k=10$ to $k=110$ (the average answer size) growing by 20 at each step.

The results shown in Figure \ref{fig:real_search} tell a different story from Figure \ref{fig:synth_search}. While $D^3L$ still outperforms $TUS$, the performance gap shrinks considerably, particularly for $k > 50$. This is because \textit{Smaller Real} contains a greater ratio of numeric values (shown in Figure \ref{fig:numerical}) and fewer tables overall than \textit{Synthetic} ($700$ v. $5000$). While $D^3L$ spends computation time in considering numeric attributes, they are completely ignored by $TUS$. Thus, the performance leaks that were significant before do not occur in this case. The flip side, for $TUS$, is a loss of about 0.2 in both precision and recall at $k=110$.

We also report the average search time of $Aurum$ obtained for 100 queries on \textit{Smaller real}, with a graph structure that accommodates a result size of at least $110$: $18.37$ seconds.

\begin{table}[t]
\centering
\small
\caption{Space overhead for different repositories.}
\begin{tabular}{|c||c|c|c|}
\hline
 & \textit{Synthetic} & \textit{Smaller Real} & \textit{Larger Real (sample)} \\ 
 \hline \hline
$D^3L$ & 69\% & 33\% & 58\%  \\ \hline 
$TUS$  & 56\% & 19\% & 32\%  \\ \hline
$Aurum$  & 55\% & 20\% & 29\%  \\ \hline
\end{tabular}
\label{tab:space}
\end{table}

\smallskip \noindent \textbf{Experiment \stepcounter{experiment}\theexperiment\xspace: Space overhead of the indexes.} 
In Table \ref{tab:space} we report the total space occupied by indexes, relative to the data lake size, for three repositories: \textit{Synthetic} (1.1 GB), \textit{Smaller Real} (600 MB), and a sample of \textit{Larger Real} (3 GB). We used a sample of the \textit{Larger Real} because building the $TUS$ indexes for the full 12 GB repository requires more than 20 hours. We also report the combined space overhead of $Aurum$'s graph data structure, profile store, and LSH indexes.

For the \textit{Synthetic} repository, $TUS$ and $Aurum$ occupy $13\%$ less space compared to $D^3L$. This is because $D^3L$ indexes four types of relatedness evidence, as opposed to only three in $TUS$ and $Aurum$. The differences in occupied space increase for \textit{Smaller} and \textit{Larger Real} repositories. This is because, in addition to creating more indexes, $D^3L$ uses finer--grained features for relatedness discovery, which results in more related attributes being discovered, which, in turn, results in more entries (buckets) per index.

\subsection{Impact of join opportunities}

We report on the impact of identifying join paths that start from some table from the top-$k$. Our stated motivation for searching join paths is to populate as many target attributes as possible. As such, we adopt notions of coverage and attribute precision to compare what is achievable when we take into account join opportunities and when we do not.

In order to define a measure for coverage, firstly, let $\mathcal{S}^k=\{S_1, \ldots S_k\}$ be the $k$-most related datasets to a given target $T$. Given a datasets $S_i \in \mathcal{S}^k$, let $\mathcal{J}_{S_i}$ be the set of all join paths $J_l$ that start from $S_i$. We denote the arity of $T$ as $arity(T)$ and the projection from $S_i$ of the attributes that are related to some attribute in $T$ by $\pi_{related(T)}(S_i)$.

We define the coverage of $S_i$ on $T$ as the ratio of attributes in $T$ that are related to some attribute in $S_i$:

\begin{equation} \label{eq:simple_cov}
\small
    cov_{S_i}(T) = \frac{arity(\pi_{related(T)}(S_i))}{arity(T)}
\end{equation}
Note that the coverage of a join path $J_l \in \mathcal{J}_{S_i}$ can be defined in a similar way by replacing $S_i$ with the result of the join. For the purpose of our comparison though, we are interested in the combined coverage of all the join path results in $\mathcal{J}_{S_i}$, since each join path can contribute with new attributes to the target. As such, we define the coverage of $\mathcal{J}_{S_i}$ on $T$ as:
\begin{equation} \label{eq:join_cov}
\small
    cov_{\mathcal{J}_{S_i}}(T) = \frac{arity(\bigcup\limits_{J_l \in \mathcal{J}_{S_i}} \pi_{related(T)}(J_l))}{arity(T)}
\end{equation}

From Equations \ref{eq:simple_cov} and \ref{eq:join_cov} we average the coverage measures of all $S_i \in \mathcal{S}^k$ and use the resulting measures, with various values for $k$, to show how the target coverage increases when we consider datasets from join paths.

For the purpose of computing attribute precision for a dataset $S_i$, we count an alignment between an attribute of $S_i$ and a target attribute as a true positive if, as per the ground truth, the two attributes are related (as defined by Definition \ref{def:relatedness}), and as a false positive when they are not related. Correspondingly, we extend this definition for computing attribute precision for a set of join paths $\mathcal{J}_{S_i}$, and, firstly, we find the set of all attributes of datasets in $\mathcal{J}_{S_i}$ that are aligned with the same target attribute, and count this set as a true positive if it contains at least one element that is related with that target attribute in the ground truth, and as a false positive otherwise. As before, for both cases, we report the average attribute precision of all $S_i$ in $\mathcal{S}^k$, at various values for $k$.

In the experiments, we use the \textit{Synthetic} and \textit{Smaller Real} repositories (for which we have the ground truth available). Our hypothesis is that by considering join paths we can identify relevant datasets that are not part of the initial ranked solution, but can improve the target coverage.

We report the target coverage and attribute precision with ($D^3L+J$/$Aurum+J$) and without ($D^3L$/$Aurum$/$TUS$) augmenting the top-$k$ result with joinable datasets. Note that the graph structure built by $Aurum$ includes $PK/FK$ candidate relationships, but $TUS$ does not address joinability discovery.

\begin{figure}[t]
\centering
\begin{subfigure}{.23\textwidth}
    \centering
    \includegraphics[width=\linewidth, trim=4 4 4 4,clip]{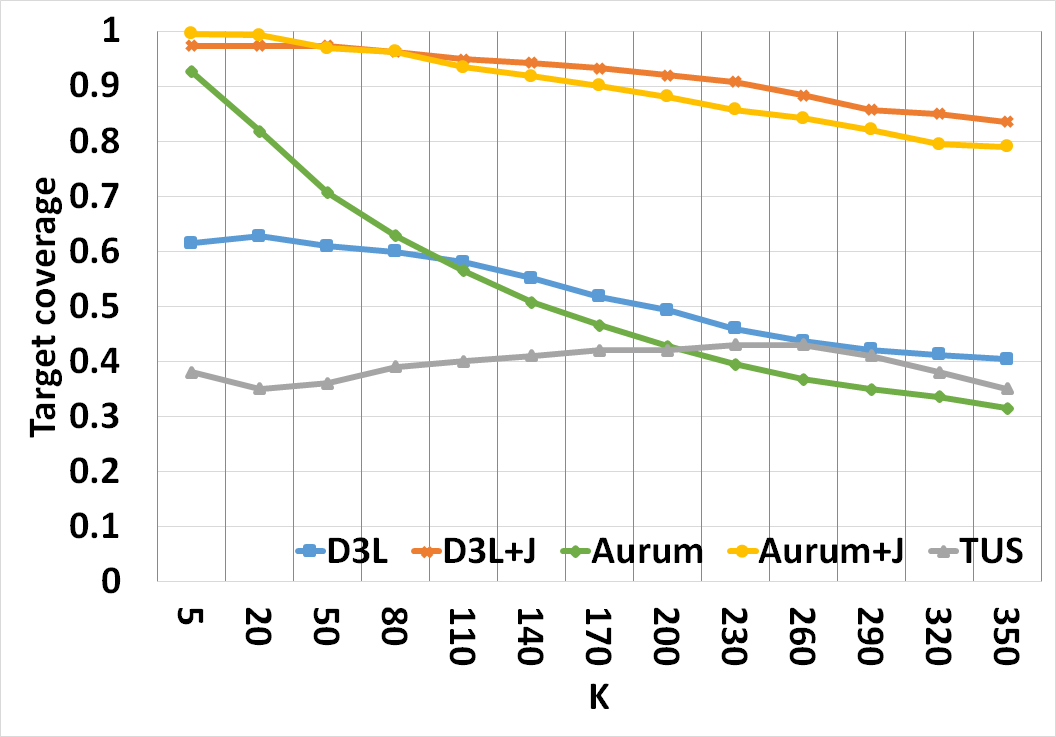}
    \caption{Target coverage}
    \label{fig:synthCov}
\end{subfigure}
\begin{subfigure}{.23\textwidth}
    \centering
    \includegraphics[width=\linewidth, trim=4 4 4 4,clip]{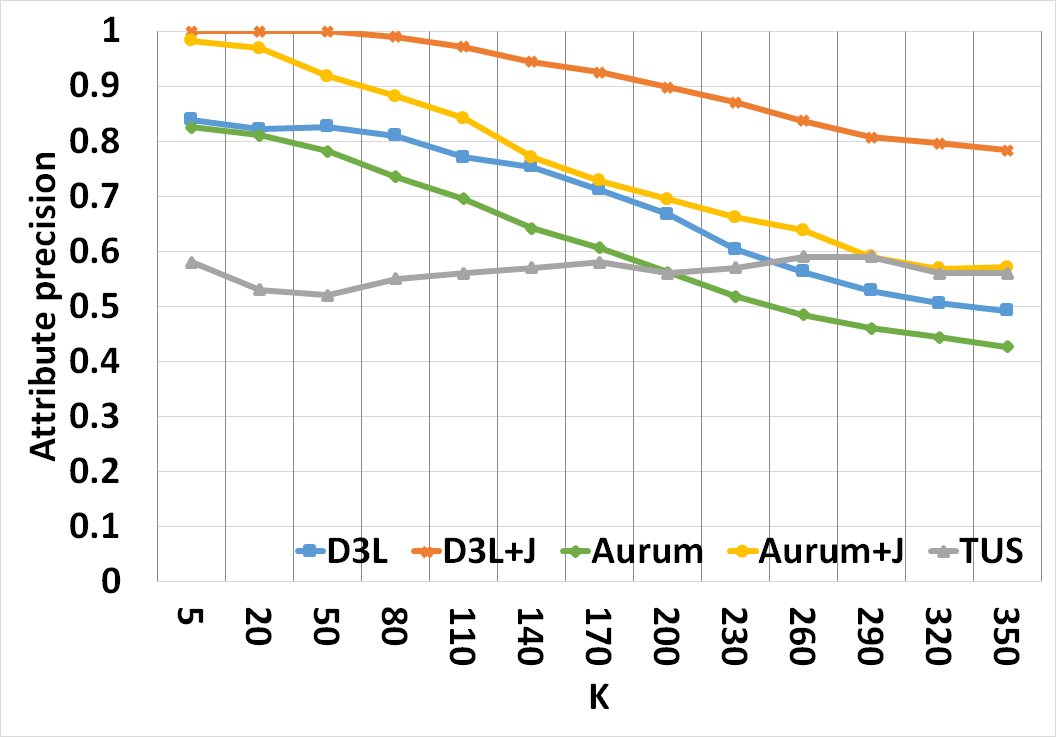}
    \caption{Attribute precision}
    \label{fig:synthAttrPrec}
\end{subfigure}
\caption{Coverage and precision on \textit{Synthetic}}
\label{fig:synthCovPrec}
\end{figure}

\smallskip \noindent \textbf{Experiment \stepcounter{experiment}\theexperiment\xspace (on \textit{Synthetic}): Target coverage as answer size grows.} 
The $D^3L+J$ and $Aurum+J$ curves from Figure \ref{fig:synthCov} suggests that the two systems are able to cover most target attributes by following join paths. The sharp decrease in coverage when join paths are not considered confirms our hypothesis that join paths allow us to identify sources potentially far away from the target but relevant for maximizing its coverage. The superior coverage manifested by $Aurum$ for $k \in [5, 80]$ can be explained by the fact that the ranking strategy employed in $Aurum$ favours the quantity of covered target attributes, over the strength of the relatedness. In $D^3L$'s case, the aggregation framework splits the ranking criteria between the number of covered attributes and the strength of the similarity. $TUS$ seems to return many unrelated datasets with the given target at the top of the ranking and, therefore, is less effective in covering it.


\smallskip \noindent \textbf{Experiment \stepcounter{experiment}\theexperiment\xspace (on \textit{Synthetic}): Attribute precision as answer size grows.} 
Figure \ref{fig:synthAttrPrec} shows how many of the attributes used to populate the target are correct in each case. Attribute precision is between $85\%$ and $100\%$ when populating the target with the attributes returned by $D^3L+J$ and $k < 260$, in contrast with $Aurum+j$, which decreases more sharply, i.e., a lower bound of $65\%$ at $k=260$. The results are consistent with the ones reported in Section \ref{sec:compEfec} and are the consequences of the same characteristics: finer--grained features that are more diagnostic and multi--evidence similarity signals considered by $D^3L(+J)$. Furthermore, the join paths in $D^3L+J$ are built on more than just uniqueness of values (as is the case for $Aurum+J$), i.e., they use subject attributes, and, therefore, they introduce fewer false positives and lead to the discovery of more related attributes. As before, $TUS$ returns more unrelated tables at the top of the ranking than $D^3L$ and $Aurm$ and, therefore, is less precise.

\begin{figure}[t]
\centering
\begin{subfigure}{.23\textwidth}
    \centering
    \includegraphics[width=\linewidth, trim=4 4 4 4,clip]{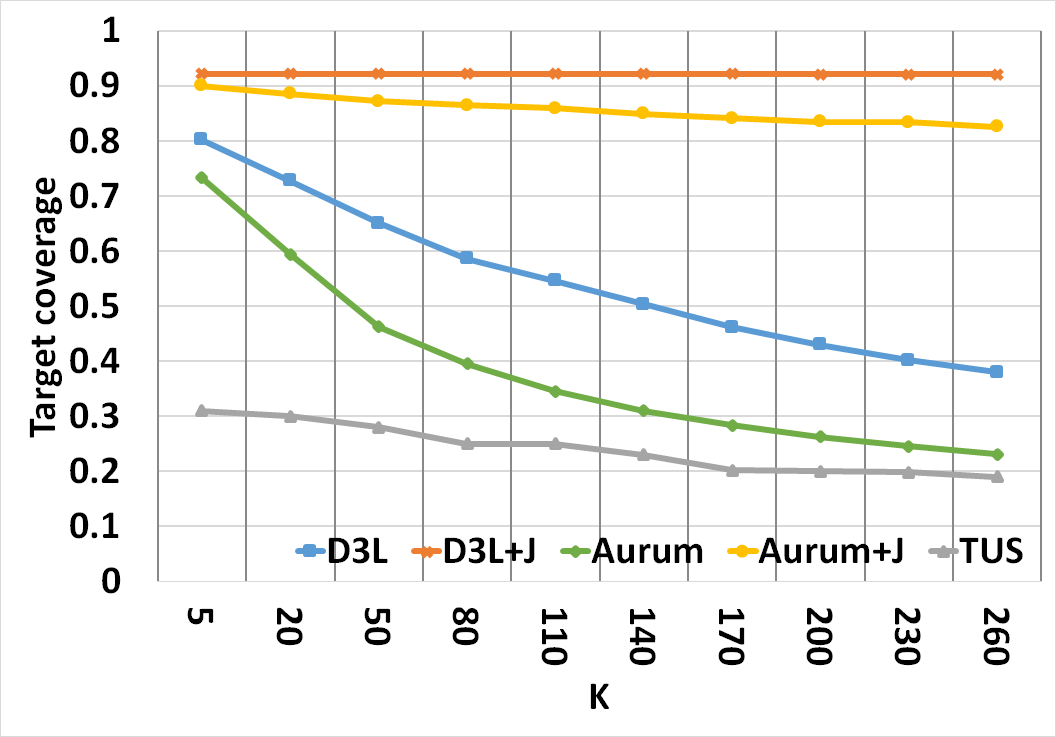}
    \caption{Target coverage}
    \label{fig:realCov}
\end{subfigure}
\begin{subfigure}{.23\textwidth}
    \centering
    \includegraphics[width=\linewidth, trim=4 4 4 4,clip]{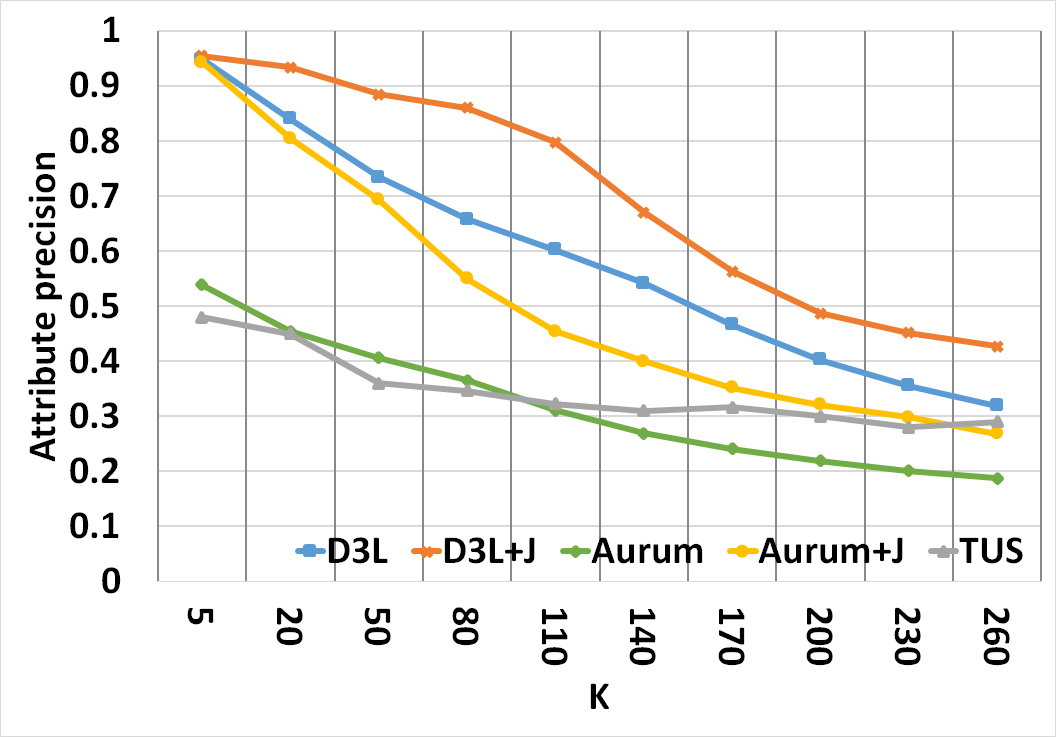}
    \caption{Attribute precision}
    \label{fig:realAttrPrec}
\end{subfigure}
\caption{Coverage and precision on \textit{Smaller Real}}
\label{fig:realCovPrec}
\end{figure}

\smallskip \noindent \textbf{Experiment \stepcounter{experiment}\theexperiment\xspace (on \textit{Smaller Real}): Target coverage as answer size grows.} 
Figure \ref{fig:realCov} shows that both $D^3L+J$ and $Aurum+J$ achieve considerable improvements in coverage over their join--unaware variants. The increase is, as expected, smaller at low $k$ values, e.g. $k \in [5,20]$, because the top of the ranking already covers the target well. As $k$ increases, the improvement in coverage becomes more significant, especially in $Aurum's$ case. This suggests, once again, that tables that are related to the target but are not included in the top-$k$ (due to an index miss, or weak relatedness signals) can be identified by traversing join paths from some top-$k$ datasets.

The low $TUS$ coverage shown in Figure \ref{fig:realCov} suggests that the top-$k$ solution covers only a small fraction of the target attributes. This is because datasets at the top of the ranking contain attributes aligned with approx. $25\%$ of target attributes, while the rest (even as $k$ increases) do not contribute  many additional attributes.

$D^3L$ proves significantly better at covering target attributes than $TUS$ and $Aurum$ for the entire interval of $k$ values. This is because, as previous experiments showed, $D^3L$ retrieves higher quality datasets (i.e., more related to the target) from the lake. The decrease of the curve as $k$ increases can be explained by the fact that the measure is an average of individual coverage values. As $k$ increases, there are more datasets with small coverage and the overall average decreases.



\smallskip \noindent \textbf{Experiment \stepcounter{experiment}\theexperiment\xspace (on \textit{Smaller Real}): Attribute precision as answer size grows.} 
Figure \ref{fig:realAttrPrec} suggests that only $35\%$ to $45\%$, and only $20\%$ to $50\%$ of the target attributes populated by $TUS$ and $Aurum$, resp., are correct. This is not surprising since the dataset--level precision reported in Section \ref{sec:compEfec} showed that at most $50\%$, in the case of $TUS$, and at most $70\%$, in the case of $Aurum$, of the retrieved datasets are indeed relevant for populating the target.

The increased precision of $D^3L$ is explained by its ability to identify attribute relatedness even when the format representation of values differs. The difference is preserved when joinable tables are considered. By including tables from the join paths in the solution, at $k \in [50, 170]$, the attribute precision increases by up to $0.2$. Note that, for $D^3L+J$, there is not much increase at the head and tail of the $k$ values interval, since datasets at the top already cover the target precisely, while datasets that are joinable with tables far away from the target provide low quality attributes. Furthermore, the precision of $D^3L+J$ does not descend below the original precision of $D^3L$, suggesting that most of the attributes contributed by the former are true positives.

Finally, as in the \textit{Synthetic} case, the use of more restrictive conditions (detailed in Section \ref{sec:joins}), i.e., the use of subject attributes, when searching for join paths compared to $Aurum+J$, allows $D^3L+J$ to cover the target more precisely.

\section{Conclusions}

We have contributed an effective and efficient solution to the problem of dataset discovery in data lakes. We have used schema-- and instance--based features to construct hash--based indexes that map the features into a uniform distance space, making it possible to take hash values similarity as relatedness measurements and, thereby, saving on computational effort.

In comparison with similar approaches from the state--of--the--art, we have empirically identified three main advantages of our proposal: (i) the use of schema and instance--level fine--grained features that are more effective in identifying relatedness, especially when similar entities are inconsistently represented; (ii) the mapping of these features to a uniform distance space that offers an aggregated view on the notion of relatedness, to which each type of similarity evidence contributes; and (iii) the discovery of join paths using LSH evidence and subject--attributes that leads to an increased and precise target coverage. These characteristics are decisive in performing more effective and more efficient dataset discovery, when compared to $TUS$, and more effective dataset discovery, at the expense of efficiency, when compared to $Aurum$, in both pragmatically generated and real--world scenarios.

\smallskip 
\noindent \textbf{Acknowledgments:} 
Work supported by the VADA Grant of the UK Engineering and Physical Sciences Research Council.

\bibliographystyle{IEEEtran}
\bibliography{main}

\begin{thebibliography}{10}
\providecommand{\url}[1]{#1}
\csname url@samestyle\endcsname
\providecommand{\newblock}{\relax}
\providecommand{\bibinfo}[2]{#2}
\providecommand{\BIBentrySTDinterwordspacing}{\spaceskip=0pt\relax}
\providecommand{\BIBentryALTinterwordstretchfactor}{4}
\providecommand{\BIBentryALTinterwordspacing}{\spaceskip=\fontdimen2\font plus
\BIBentryALTinterwordstretchfactor\fontdimen3\font minus
  \fontdimen4\font\relax}
\providecommand{\BIBforeignlanguage}[2]{{%
\expandafter\ifx\csname l@#1\endcsname\relax
\typeout{** WARNING: IEEEtran.bst: No hyphenation pattern has been}%
\typeout{** loaded for the language `#1'. Using the pattern for}%
\typeout{** the default language instead.}%
\else
\language=\csname l@#1\endcsname
\fi
#2}}
\providecommand{\BIBdecl}{\relax}
\BIBdecl

\bibitem{Furche-16}
T.~Furche, G.~Gottlob, L.~Libkin, G.~Orsi, and N.~W. Paton, ``Data wrangling
  for big data: Challenges and opportunities,'' in \emph{EDBT}, 2016.

\bibitem{Koehler-17}
M.~Koehler, A.~Bogatu, C.~Civili, N.~Konstantinou, E.~Abel, A.~A.~A. Fernandes,
  J.~A. Keane, L.~Libkin, and N.~W. Paton, ``Data context informed data
  wrangling,'' in \emph{IEEE Big Data}, 2017.

\bibitem{Konstantinou-17}
N.~Konstantinou, M.~Koehler, E.~Abel, C.~Civili, B.~Neumayr, E.~Sallinger,
  A.~A.~A. Fernandes, G.~Gottlob, J.~Keane, L.~Libkin, and N.~W. Paton, ``The
  {VADA} architecture for cost-effective data wrangling,'' in \emph{SIGMOD},
  2017.

\bibitem{Rahm-01}
E.~Rahm and P.~A. Bernstein, ``A survey of approaches to automatic schema
  matching,'' \emph{The VLDB Journal}, vol.~10, no.~4, 2001.

\bibitem{Bogatu-19}
A.~Bogatu, A.~A.~A. Fernandes, N.~W. Paton, and N.~Konstantinou, ``Synthedit:
  Format transformations by example using edit operations,'' in \emph{EDBT},
  2019.

\bibitem{Mecca-09}
G.~Mecca, P.~Papotti, and S.~Raunich, ``Core schema mappings,'' in
  \emph{SIGMOD}, 2009.

\bibitem{Mazilu19}
L.~Mazilu, N.~W. Paton, F.~A.A.A., and M.~Koehler, ``Dynamap: Schema mapping
  generation in the wild,'' in \emph{SSDBM}, 2019.

\bibitem{Indyk-1998}
P.~Indyk and R.~Motwani, ``Approximate nearest neighbors: Towards removing the
  curse of dimensionality,'' in \emph{STOC}, 1998.

\bibitem{Fernandez-18}
R.~C. Fernandez, Z.~Abedjan, F.~Koko, G.~Yuan, S.~Madden, and M.~Stonebraker,
  ``Aurum: {A} data discovery system,'' in \emph{ICDE}, 2018.

\bibitem{Nargesian-2018}
F.~Nargesian, E.~Zhu, K.~Q. Pu, and R.~J. Miller, ``Table union search on open
  data,'' \emph{PVLDB}, vol.~11, no.~7, Mar. 2018.

\bibitem{Cafarella-08}
M.~J. Cafarella, A.~Y. Halevy, D.~Z. Wang, E.~Wu, and Y.~Zhang, ``Webtables:
  exploring the power of tables on the web,'' \emph{PVLDB}, 2008.

\bibitem{Elmeleegy-09}
H.~Elmeleegy, J.~Madhavan, and A.~Y. Halevy, ``Harvesting relational tables
  from lists on the web,'' \emph{The VLDB Journal}, vol.~2, no.~1, 2009.

\bibitem{Lehmberg-17}
O.~Lehmberg and C.~Bizer, ``Stitching web tables for improving matching
  quality,'' \emph{PVLDB}, vol.~10, no.~11, 2017.

\bibitem{Ling-13}
X.~Ling, A.~Y. Halevy, F.~Wu, and C.~Yu, ``Synthesizing union tables from the
  web,'' in \emph{IJCAI}, 2013.

\bibitem{DasSarma-2012}
A.~Das~Sarma, L.~Fang, N.~Gupta, A.~Y. Halevy, H.~Lee, F.~Wu, R.~Xin, and
  C.~Yu, ``Finding related tables,'' in \emph{SIGMOD}, 2012.

\bibitem{Fernandez18_sem}
R.~Fernandez, E.~Mansour, E.~Qahtan, A.~Elmagarmid, I.~Ilyas, S.~Madden,
  M.~Ouzzani, M.~Stonebraker, and N.~Tang, ``Seeping semantics: Linking
  datasets using word embeddings for data discovery,'' in \emph{ICDE}, 2018.

\bibitem{Datar-04}
M.~Datar, N.~Immorlica, P.~Indyk, and V.~S. Mirrokni, ``Locality-sensitive
  hashing scheme based on p-stable distributions,'' in \emph{SoCG}, 2004.

\bibitem{Broder-1997}
A.~Broder, ``On the resemblance and containment of documents,'' in
  \emph{SEQUENCES}, 1997.

\bibitem{Charikar-2002}
M.~S. Charikar, ``Similarity estimation techniques from rounding algorithms,''
  in \emph{STOC}, 2002.

\bibitem{Bawa-05}
M.~Bawa, T.~Condie, and P.~Ganesan, ``{LSH} forest: self-tuning indexes for
  similarity search,'' in \emph{WWW}, 2005.

\bibitem{Zhu-2016}
E.~Zhu, F.~Nargesian, K.~Q. Pu, and R.~J. Miller, ``Lsh ensemble:
  Internet-scale domain search,'' \emph{PVLDB}, vol.~9, no.~12, Aug. 2016.

\bibitem{Miller-18}
R.~Miller, ``Open data integration,'' \emph{PVLDB}, vol.~11, no.~12, 2018.

\bibitem{Cafarella-09}
M.~J. Cafarella, A.~Y. Halevy, and N.~Khoussainova, ``Data integration for the
  relational web,'' \emph{PVLDB}, vol.~2, no.~1, 2009.

\bibitem{Terrizzano15}
I.~Terrizzano, P.~Schwarz, M.~Roth, and J.~Colino, ``Data wrangling: The
  challenging yourney from the wild to the lake,'' in \emph{CIDR}, 2015.

\bibitem{Halevy16}
A.~Halevy, F.~Korn, N.~F. Noy, C.~Olston, N.~Polyzotis, S.~Roy, and S.~E.
  Whang, ``Goods: Organizing google's datasets,'' in \emph{SIGMOD}, 2016.

\bibitem{Mikolov-13}
T.~Mikolov, I.~Sutskever, K.~Chen, G.~Corrado, and J.~Dean, ``Distributed
  representations of words and phrases and their compositionality,'' in
  \emph{NIPS}, 2013.

\bibitem{Grave-17}
E.~Grave, T.~Mikolov, A.~Joulin, and P.~Bojanowski, ``Bag of tricks for
  efficient text classification,'' in \emph{EACL}, 2017.

\bibitem{Stats-99}
W.~J. Conover, \emph{Practical nonparametric statistics}.\hskip 1em plus 0.5em
  minus 0.4em\relax Wiley, 1999.

\bibitem{Venetis-2011}
P.~Venetis, A.~Y. Halevy, J.~Madhavan, M.~Pa\c{s}ca, W.~Shen, F.~Wu, G.~Miao,
  and C.~Wu, ``Recovering semantics of tables on the web,'' \emph{PVLDB},
  vol.~4, no.~9, Jun. 2011.

\bibitem{Hsieh-08}
C.~Hsieh, K.~Chang, C.~Lin, S.~S. Keerthi, and S.~Sundararajan, ``A dual
  coordinate descent method for large-scale linear {SVM},'' in \emph{ICML},
  2008.

\bibitem{Papenbrock-15}
T.~Papenbrock, T.~Bergmann, M.~Finke, J.~Zwiener, and F.~Naumann, ``Data
  profiling with metanome,'' \emph{PVLDB}, vol.~8, no.~12, 2015.

\bibitem{Suchanek-07}
F.~M. Suchanek, G.~Kasneci, and G.~Weikum, ``Yago: A core of semantic
  knowledge,'' in \emph{WWW}, 2007.

\end{thebibliography}

\end{document}